\documentclass{WileyMSP-template}

\usepackage{amsmath}
\usepackage{hyperref}
\usepackage{multirow, booktabs}
\usepackage{float}
\usepackage{subcaption}

\usepackage{tikz}
\usepackage{lipsum}    
\usepackage{siunitx}

\usepackage{color,soul}

\usepackage{adjustbox}
\usepackage{xcolor}
\usepackage{tabularray}
\definecolor{tableheader}{HTML}{FF6361}
\usepackage{listings}
\usepackage{multibib}
\usepackage{gensymb}

\begin{document}

\pagestyle{fancy}
\rhead{\includegraphics[width=2.5cm]{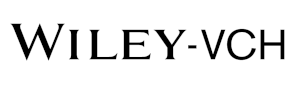}}

\title{Long Spin Relaxation Times in CVD-Grown Nanodiamonds}

\maketitle


\author{Jeroen Prooth}
\author{Michael Petrov}
\author{Alevtina Shmakova}
\author{Michal Gulka}
\author{Petr Cigler}
\author{Jan D'Haen}
\author{Hans-Gerd Boyen}
\author{Milos Nesladek*}


\dedication{}

\begin{affiliations}
J. Prooth$^{1*}$, M. Petrov$^{1*}$, A. Shmakova$^1$, Dr. M. Gulka$^2$, Prof. Dr. P. Cigler$^2$, Prof. Dr. J. D'Haen$^1$, Prof. Dr. H.-G. Boyen$^1$, Prof. Dr. M. Nesladek$^1$\\
$^1$Institute for Materials Research (IMO), Hasselt University, Wetenschapspark 1, B-3590, Diepenbeek, Belgium\\
$^2$Institute of Organic Chemistry and Biochemistry of the CAS, Prague, 166 10, Czech Republic\\
$^*$Authors contributed equally\\

Email Address: milos.nesladek@uhasselt.be

\end{affiliations}



\begin{abstract}

Currently, the primary applications of fluorescent nanodiamonds (FNDs) are in the area of biosensing, by using photoluminescence or spin properties of colour centres, mainly represented by the Nitrogen Vacancy (NV) point defect. The sensitivity of NV-FNDs to external fields is, however, limited by crystallographic defects, which influence their key quantum state characteristics - the spin longitudinal (\textit{\textit{T$_1$}}) and spin transversal (\textit{T$_2$}) relaxation and coherence times, respectively. We report on utilising an advanced FND growth technique consisting of heterogeneous nucleation on pre-engineered sites to create FNDs averaging around 60 nm in size, with mean longitudinal coherence times of 800 $\mu$s and a maximum over 1.8 ms, close to bulk theoretical values. This is a major, nearly ten-fold improvement over commercially available nanodiamonds for the same size range of 50 to 150 nm. Heavy-N doped nanodiamond shells, important for sensing events in nm proximity to the diamond surface, are fabricated and discussed in terms of re-nucleation and twinning on \{111\} crystal facets. We also discuss scalability issues in order to enable the production of FND volumes matching the needs of sensing applications.

\end{abstract}


\section{Introduction}

Nanodiamond (ND) is a fascinating class of nano-material because of its unique properties, such as low cytotoxicity and biocompatibility.\cite{https://doi.org/10.1002/advs.202200059, doi:10.1021/acssensors.1c00415, doi:10.1021/acs.analchem.1c04536} NDs can be used for nanoscale sensing based on fluorescent properties of incorporated point (colour) defect centres. For these purposes, this type of nanodiamond is referred to as Fluorescent Nanodiamond (FND). 
Further on, the paramagnetic, e.g. spin properties of colour centres, enable using FNDs for sensing the magnetic field noise close to the ND surface, leading to novel measuring concepts such as nano-NMR.\cite{PhysRevLett.108.197601, Rondin_2014, PhysRevLett.99.250601, doi:10.1126/science.1231675, Glenn2018} Most of the FNDs sensors are based on
the nitrogen-vacancy (NV) centre, that has attracted significant attention due to its bright photoluminescence and negligible photobleaching.\cite{Jin-Xu} The NV centre consists of a nitrogen atom next to a vacancy and has been studied fundamentally and used in initial applications. The NV spin ground triplet state can easily be manipulated by microwaves at ambient conditions. The crucial challenge for developing nanoscale sensing probes, based on the NV colour centres, is to engineer FNDs with high spin relaxation (\textit{T$_1$}) and spin coherence (\textit{T$_2$}) times, which differ significantly for various types of FNDs, and which limit a wide spread use of FNDs sensors in biology and chemistry fields, for example in drug delivery, bio-imaging, and bio-nanosensors for applications operating at room temperatures. Currently, commercial FNDs have \textit{T$_1$} times in the range of 100 microseconds, while the bulk diamond \textit{T$_1$} limit is about three milliseconds, thus limiting FND use significantly.\cite{Ermakova} Here, we present data on about 8-fold improvement of the \textit{T$_1$} time by Chemical Vapor Deposition (CVD) technique, supported by understanding the underlying physics.
\hfill \break

\begin{figure}
\centering 
    \includegraphics[width=0.7\columnwidth]{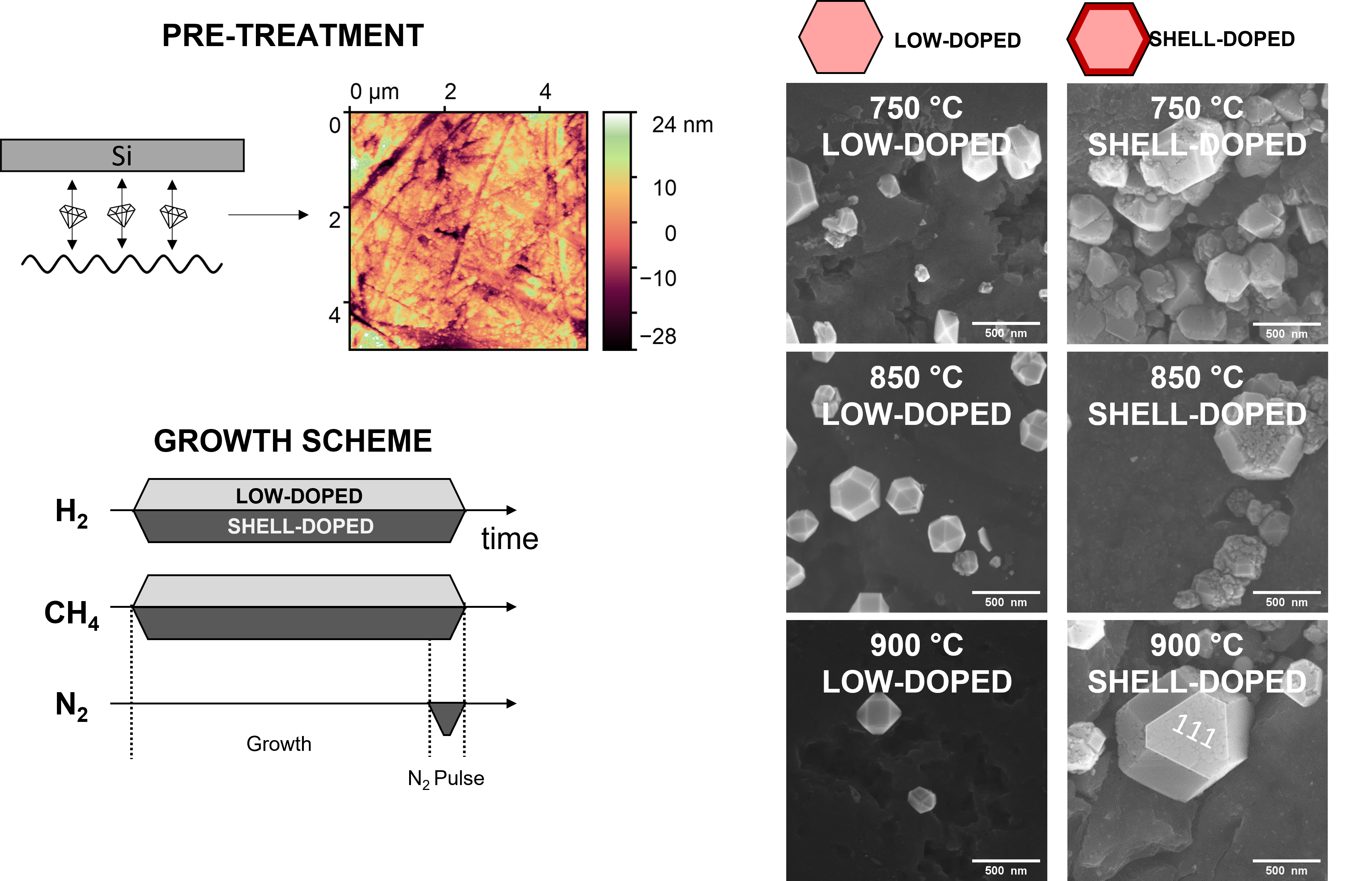}
    \captionof{figure}{A nano-roughening pre-treatment on silicon is performed by using an ultrasonic vibration table on which diamond micro-particles are spread and on which the substrate is vibrated. This produces nano-roughness, and leaves carbon traces on the substrate surface,  pictured in the AFM image. The growth scheme shows the introduction of gasses during the growth stage. As a base diamond material we use low N-doped samples, whereas for the shell-doped particles we created in addition a highly nitrogen-doped shell. Images of particles grown at 750 °C, 850 °C and 900 °C taken with a GeminiSEM 450 at 5 kV. Left side shows low-doped samples, with nitrogen shell-doped samples on the right. Shell-doped particles show preferential re-nucleation on their \{111\} facets due to the too-high nitrogen concentration, whereas low-doped particles show no such structures. Twinning can be found for all conditions.}
    \label{fig:Morphology}
\end{figure}

Currently, commercial FNDs are mainly prepared from high pressure, high temperature (HPHT) diamond by milling \cite{Boudou_2009} and subsequent irradiation and annealing. HPHT FNDs have a typical NV \textit{T$_1$} relaxation time between 50 and 100 µs.\cite{REDOXHPHT} As the sensitivity of NV centre to magnetic fields is limited by the \textit{T$_1$} relaxation time, the ability to engineer FNDs with higher \textit{T$_1$} times would improve the detection sensitivity.\cite{deGuillebon2020} Nevertheless, this requires precise control over particle properties such as size, morphology, defects, and dopants. The HPHT FND approach, based on milling and grinding N-containing bulk diamond, produces FNDs that range in size from around 10 nm to 1 µm. This mechanical process makes these FNDs highly irregular and jagged, generating a significant amount of spin-active surface defects, dangling bonds and subsurface damage.\cite{Schirhagl2014} Further processing using high energy particle irradiation, employed to create nitrogen-vacancies, can cause internal lattice damage and introduce spin active defects, strain fields, or vacancy clusters. For HPHT nanodiamonds, one usually irradiates them with electrons (e$^-$), protons (p$^+$), or He$^{2+}$ and Li$^+$ ions. An annealing step allows the vacancies to diffuse towards the substitutional nitrogen and create NV centres, while at the same time some of the damage can be repaired. Nevertheless, a study done on several HPHT nanodiamonds irradiated by these methods showed that particles with sizes around 50 to 150 nm had \textit{T$_1$} times between 50 and 100 µs for those irradiated with e$^-$, and He$^{2+}$ and up to 150 µs for particles irradiated with p$^+$.\cite{Reineck} Other studies have shown similar results, with \textit{T$_1$} times ranging around 100 µs.\cite{REDOXHPHT} FNDs with a size of around 15 nm showed lower \textit{T$_1$} times around 25 µs.\cite{Ermakova} Another method for large-scale production of fluorescent nanodiamonds (FND) using neutron irradiation is capable of producing significant amounts of nanodiamonds, typically on the order of around 100 grams.\cite{Havlik} However, the FNDs' obtained \textit{T$_1$} times are short compared to their expected bulk values. In another work, the milling technique has been improved, leading to 1.2 ms \textit{T$_1$} times, however, only a few percent of FNDs contained NV centres.\cite{Knowles2014}  Chemical vapour deposition (CVD) using isotopically pure $^{12}$C as a source gas was used to grow diamond films, which were then again milled to produce FNDs.\cite{PhysRevB.105.205401} Using this technique \textit{T$_2$} coherence close to 400 microseconds was reported. However, all methods employing milling face challenges related to defect formation, chipping, and the production of irregular sizes caused by the milling process. One of the techniques to eliminate milling is based on the etching of diamond to form diamond nano-pillars through the use of nano-processing and nano-lithography.\cite{Trusheim} This approach results in nearly bulk-like diamond \textit{T$_2$} spin coherence times of approximately 210 microseconds. However, it should be noted that this technique produces nanopillars rather than round nanodiamonds, which may not be suitable for certain applications such as cellular biology, where nano-pillars may pose risks of piercing the cell membrane.
\hfill \break
 
A FND preparation technique, pursued here, is the direct nanocrystal growth using CVD. This means that instead of growing microcrystals and milling them down to nanoparticles in a top-down approach, we grow them directly on the substrate in a bottom-up way, avoiding surface defects and shape irregularities which are introduced through the milling process.
To initiate the growth of nanodiamonds, a seed with a diamond structure is required,\cite{D1RA00397F, Gebbie2018} such as detonation nanodiamond (DND). However, this technique would quickly result in the formation of a thin film. In particular, for achieving high \textit{T$_1$} times it is necessary that the crystals grow individually and develop clear facets. In that sense, the nanocrystalline nature would reduce the \textit{T$_1$} times. Alternatively, diamondoids can initiate nucleation; for example, pentamantane can be used for seeding.\cite{Gebbie2018, Tzeng-VerticalSub, penta} This requires bonding of diamondoids to the substrate, because diamondoids easily sublimate at temperatures lower than the diamond growth temperature,\cite{ishiwata2015molecular} limiting diamond nucleation. In all these cases, mainly, a continuous nanodiamond film is formed. In this paper, we adopt a different approach, one that is both simple and easy to perform i.e. heterogeneous nucleation. Spontaneous heterogeneous nucleation, meaning nucleation on substrates chemically or crystallographically different from the material to be grown, has been employed before, but the resulting nucleation densities are extremely low.\cite{Gebbie2018} Therefore, to enhance it, one can perform a pre-treatment such as mechanical nano-roughening.\cite{NESLADEK2006607} The advantage of these methods is that no seed is required, and one has, in theory, complete control over particle properties from the start of the growth.
\hfill \break

This technique of synthesizing nanodiamonds through heterogeneous nucleation results in increased \textit{T$_1$} times of around 800 µs for particles averaging 60 nm, which is an 8 fold improvement compared to HPHT FNDs, where \textit{T$_1$} reaches roughly 100 µs for similar-sized particles. The maximum \textit{T$_1$} times exceed 1.8 ms, which is close to bulk diamond values. For this study we have performed measurements over several growth conditions, and the \textit{T$_1$} is consistent between samples.
\hfill \break

\indent Since the NVs are naturally generated during the CVD process, consequently, the potential deterioration of NV spin coherence properties is limited, compared to when implantation is used. However \textit{T$_2$} times will be influenced by the spin bath induced by paramagnetic defects generated during growth, or by surface defects. \cite{ PhysRevLett.108.197601, PhysRevB.105.205401, Balasubramanian} In the CVD process, the NV yield is rather low, and a large portion of N forms paramagnetic substitutional defects. Therefore, in this work, we also evaluate NV formation efficiency.\\
\indent Further on, we address the following questions related to achieving the highest \textit{T$_1$} times: We study FND growth at different conditions, such as the substrate temperature, and we investigate the \textit{T$_1$} value's dependence on the nanodiamond size. Due to the interest in biological applications and the need to have the NV centre in nm depth from the FND surface for increased sensitivity to the environment, we have developed an N$_2$-pulse doping technique that allows us to increase the N$_2$ concentration in the gas phase at the very end of diamond growth, producing highly nitrogen-doped FND shells. We studied \textit{T$_1$} in such samples and compared them with FND samples without the shell. We find that re-nucleation and twinning are present due to high N-concentration in the gas phase, and defects generated in this step can still limit achieving higher \textit{T$_1$} times.

\section{Results and Discussion}
\subsection*{Nanodiamond Growth}
To nucleate diamond nanoparticles with high density, we perform nano-roughening followed by an optimised and specifically designed CVD process, \textbf{Figure} \ref{fig:Morphology}. We grow two sets of samples: reference particles (low-doped), which have no additional nitrogen added, and shell-doped particles, which have a nitrogen pulse applied at the final step of their growth. To grow well-separated individual nanocrystals rather than films, we have optimised our pretreatment method, allowing us to tune the ND coverage on the substrate by modifying the roughening duration. With this process, we can produce well-faceted nanodiamond crystals, \textbf{Figure} \ref{fig:ND-Morphology}. In the Supplementary Information, one can find an additional set of SEM images, showing ND crystals of small sizes, however taking sharp images on non-conductive particles is challenging. We therefore also tried to deposit a Pd-Au coating to reduce charging effects.
\hfill \break

Adjusting the duration of the ultrasonic treatment allows us to tune the nucleation density and produce individual crystals at a higher density. Assuming a roughly 30\% coverage, a particle diameter of 50 nm, a growth rate of 1 micrometer per hour, and a 15 cm substrate, it is possible to obtain about 2.6 mg of NDs using such a system. As illustrated in \textbf{Figure} \ref{fig:SEM-PL-SizeDistr}, at a temperature of 700 °C, roughly 35\% of NDs fall within the range of 30-60 nm, which would allow us to get roughly 1 mg of these particles within a single run, enough for further chemical processing.
\hfill \break

In this work we do not concentrate on achieving high quantities and grow on small, 5 x 5 mm$^2$ silicon substrates. Removal of NDs from the wafer can be done through etching, however this process requires the use of HF acid, which is undesirable. Another approach is to use a thin metallic intermediate layer,\cite{DEGUTIS2016163} on which NDs can be grown. The wet etching of such a layer is instantaneous as it can be very thin, 50 nm or less, making the process efficient. Using intermediate layers, NDs can be deposited on various substrates such as quartz, sapphire, and others, allowing for reusability. We perform preliminary measurements on NDs grown on an intermediate molybdenum layer (see Supporting Information) to demonstrate the feasibility of the proposed methods, however most of the work is done on particles grown on silicon substrates. We have measured \textit{T$_1$} of nanodiamonds grown on molybdenum inter-layers and achieve similar \textit{T$_1$} as those on silicon, see Supporting Information.

\begin{figure}
\centering 
    \includegraphics[width=0.3\columnwidth]{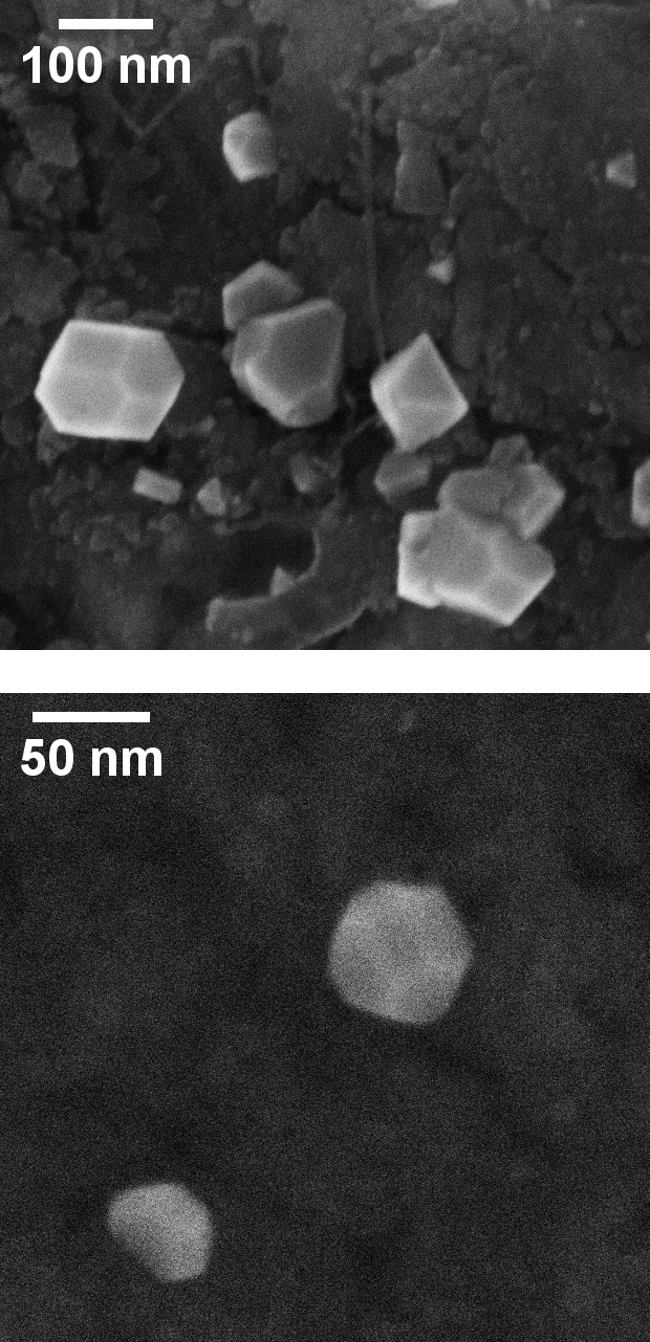}
    \captionof{figure}{SEM images showing typical particle morphology for our growth conditions. Particles shown here do not have a $\delta$-doped shell. Additional images can be found in the Supplementary Information.}
    \label{fig:ND-Morphology}
\end{figure}

\subsection*{Size and \textit{T$_1$} distributions}
Using a combination of scanning electron microscopy (SEM) and photoluminescence mapping (PL), we are able to determine the size of our particles and the average amount of NVs they contain and therefore find important parameters such as the product of the N incorporation rate and the NV generation yield, as discussed further. We match PL distributions to size distributions obtained with SEM using the shape of the curves. With this we find the correspondence between size and photoluminescence for a given laser power. This correspondence is further used to determine sizes of specifically those particles which we measure \textit{T$_1$} for.
\hfill \break

For this evaluation we use both low-doped FNDs and FNDs equipped with a 10 nm thick highly N-doped shell, as discussed in the experimental section. The size distributions of the nanodiamonds grown at various temperatures were first determined using SEM. Further on, by identifying the FNDs in the PL map and knowing the PL rate per NV, we correlate the size of particles to their \textit{T$_1$} time.
The SEM photos are analyzed with ImageJ and Python, while the PL images are analyzed with a custom-made Python program that uses the SciPy package. Due to thresholding and blurring, we estimate our inaccuracy on the diameter to be around 15 nm, or two pixels, given by the used SEM magnification. This error results from either overestimating or underestimating the border of our nanodiamonds. The Experimental Section and Supporting Information provide a detailed description of the complete workflow.
\hfill \break

Figure \ref{fig:SEM-PL-SizeDistr}a, for the NV shell doped samples, and Figure \ref{fig:SEM-PL-SizeDistr}b, for the low-doped samples, both display the FND size distributions. The computed size distributions from PL are shown in blue, and SEM distributions are shown in red. The diameter was determined from the PL intensity using the formula below; for the derivation, see the Supporting Information:

\begin{equation}
\centering
d = \sqrt[3]{\frac{3I_\text{tot}a^3}{4\pi \frac{[\text{N}_2]}{[\text{CH}_4]} I_\text{s} N_{i} N_{c}}}
\label{eq:PL-Calc-Size}
\end{equation}

with \textit{d} being the diameter, \textit{$I_{tot}$} the total measured photoluminescence of a single diamond particle, \textit{a} the diamond unit cell size equal to 3.58 \si{\angstrom}, [N$_2$]/[CH$_4$] the nitrogen to methane ratio, \textit{$I_s$} the single NV photoluminescence under constant laser power, \textit{N$_{i}$} the nitrogen incorporation rate, and \textit{N$_{c}$} the N to NV conversion rate.
The FNDs grown at 700 °C exhibit the narrowest size distribution in Figure \ref{fig:SEM-PL-SizeDistr}a, with a mean of about 58 nm. The mean size increases to higher values as the temperature is raised, and in some conditions, a secondary peak can be seen for larger particle sizes. During our image analysis, we employ a technique known as “water-shedding”, which is a powerful image analysis technique for segmenting coupled particles and detecting their boundaries.\cite{WaterSheddingPaper} As a result, this peak cannot be attributed to several clustered FNDs mistaken for one particle.
In addition, SEM images visually display a collection of smaller particles, see the Supporting Information. Due to the nature of sample pre-treatment (see the Experimental Section) we anticipate that some of the diamond debris from the roughening is still left on the silicon substrate, even after the cleaning step, and behaves as seeds. We expect that the size of this debris exceeds the critical nucleation size, allowing it to continue growing throughout the CVD process and causing a distribution of larger sizes (2$^{nd}$ peak in the distribution).
We attribute the peak on the left, for FND of 50–60 nm in size, to heterogeneous nucleation, i.e., sp$^3$ carbon residua, which transform to nuclei after an incubation period.\cite{https://doi.org/10.1002/adma.200802305} Pre-nuclei compete to grow or are etched until they eventually surpass the critical nucleation size, at which point they can continue to expand. We surmise that the occasional lack in the peak division between primary and secondary ones is due to the cleaning process' reproducibility following nano-roughening, i.e removal of nanodiamond debris. \textbf{Table S2} provides a detailed description of the size distributions' mean, 75th, and 99th percentiles, meaning 75 \% and 99 \% of particles are below those threshold respectively.
\hfill \break

\textbf{Equation \ref{eq:PL-Calc-Size}} is used to fit the PL distribution, shown in Figure \ref{fig:SEM-PL-SizeDistr}a-b, to the SEM distribution. To match the PL conditions used for \textit{T$_1$} measurements, measurements were performed at 50 µW of laser power. {For these small laser powers, the PL distribution onset (blue) starts at a larger FND size with respect to the SEM detected distribution onset (red), Figure \ref{fig:SEM-PL-SizeDistr}. This is due to the low PL luminescence counts related to small FNDs. We could also access the first peak in the distribution spectra by using higher laser powers, confirming that both the PL and SEM analyses are well matched (see the Supporting Information for more details). The same [N$_2$]/[CH$_4$] and \textit{I$_s$}  are used for shell-doped samples grown at various temperatures. As the \textit{N$_i$} x \textit{N$_c$} product is the only unknown variable in Equation \ref{eq:PL-Calc-Size}, it becomes the fitting parameter for the PL distribution. We find the value of the product \textit{N$_i$} x \textit{N$_c$} for different temperatures by matching the PL distribution (Figure \ref{fig:SEM-PL-SizeDistr}a-b-blue), to the SEM distribution (Figure \ref{fig:SEM-PL-SizeDistr}a-b-red). This approach allows us to determine the NV$^-$ concentration at different temperatures, Figure \ref{fig:SEM-PL-SizeDistr}c. We show that NV formation increases six times at 700 °C compared to 850 °C. A similar result was obtained previously by Tallaire et al.,\cite{TALLAIRE} where the NV concentration for bulk growth at 780 °C increased by three times compared to 880 °C. However, in their experiment, the growth was performed exclusively on the \{100\} facet, and here, interestingly, we have a mixture of both \{100\} and \{111\} facets. Since nitrogen incorporation is more efficient in \{111\} facet,\cite{Edmons2012} we expect that it is the N-incorporation in the \{111\} facets that determines the final N-content in the case of our nanodiamond growth as compared to \cite{TALLAIRE}. We also note that the fitting values, used in Figure \ref{fig:SEM-PL-SizeDistr}, obtained for the FNDs with nitrogen shell-doped layer, which we discuss in detail below, were very similar to the corresponding (i.e., same temperature) low-doped samples, i.e similar N$_i$ x N$_c$ for similar temperatures). The fact that the $\delta$-doped shell only has a small influence on the total luminescence is discussed later in the text. As seen in Figure \ref{fig:SEM-PL-SizeDistr}b, the distributions for both cases match well, thereby also supporting our theoretical model for size distribution calculation. \textbf{Figure S15} also shows the calculated diameter of a particle compared to its measured SEM size, having only a small error of a calculated size of roughly 280 nm to a measured size of 306 nm. 
\hfill \break

\begin{figure}
    \centering 
    \includegraphics[width=0.85\linewidth]{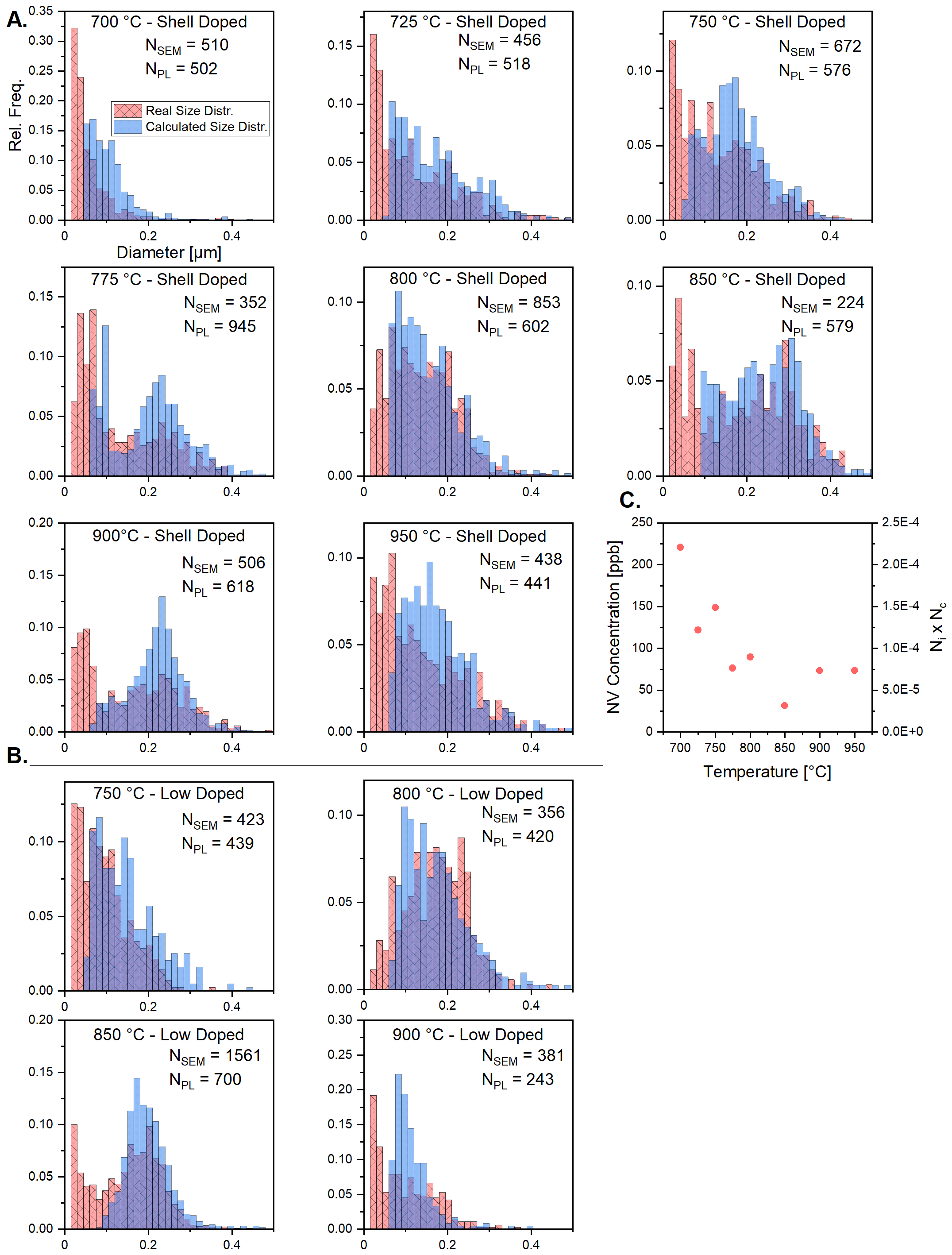}
    \captionof{figure}{A. Overview of the size distributions by SEM (red) and by calculation from PL distribution (blue) for shell-doped samples. B.  Similarly calculated distributions for low-doped samples. The calculations for corresponding temperatures used the same reactor impurity, single NV count and N$_i$ x N$_c$ rate. C. NV concentration and the product of nitrogen incorporation $N_i$ and conversion rate $N_c$ as a function of temperature showing a 6.25 fold increase at 700 °C as compared to 850 °C. N$_{SEM}$ and N$_{PL}$ are the number of particles measured for the SEM and PL distributions respectively. Particles smaller than 70 nm are absent in the PL distribution, since the sample images used for the distribution analysis are obtained at low laser power.}
    \label{fig:SEM-PL-SizeDistr}
\end{figure}

\begin{figure}
\centering 
    \includegraphics[width=0.5\columnwidth]{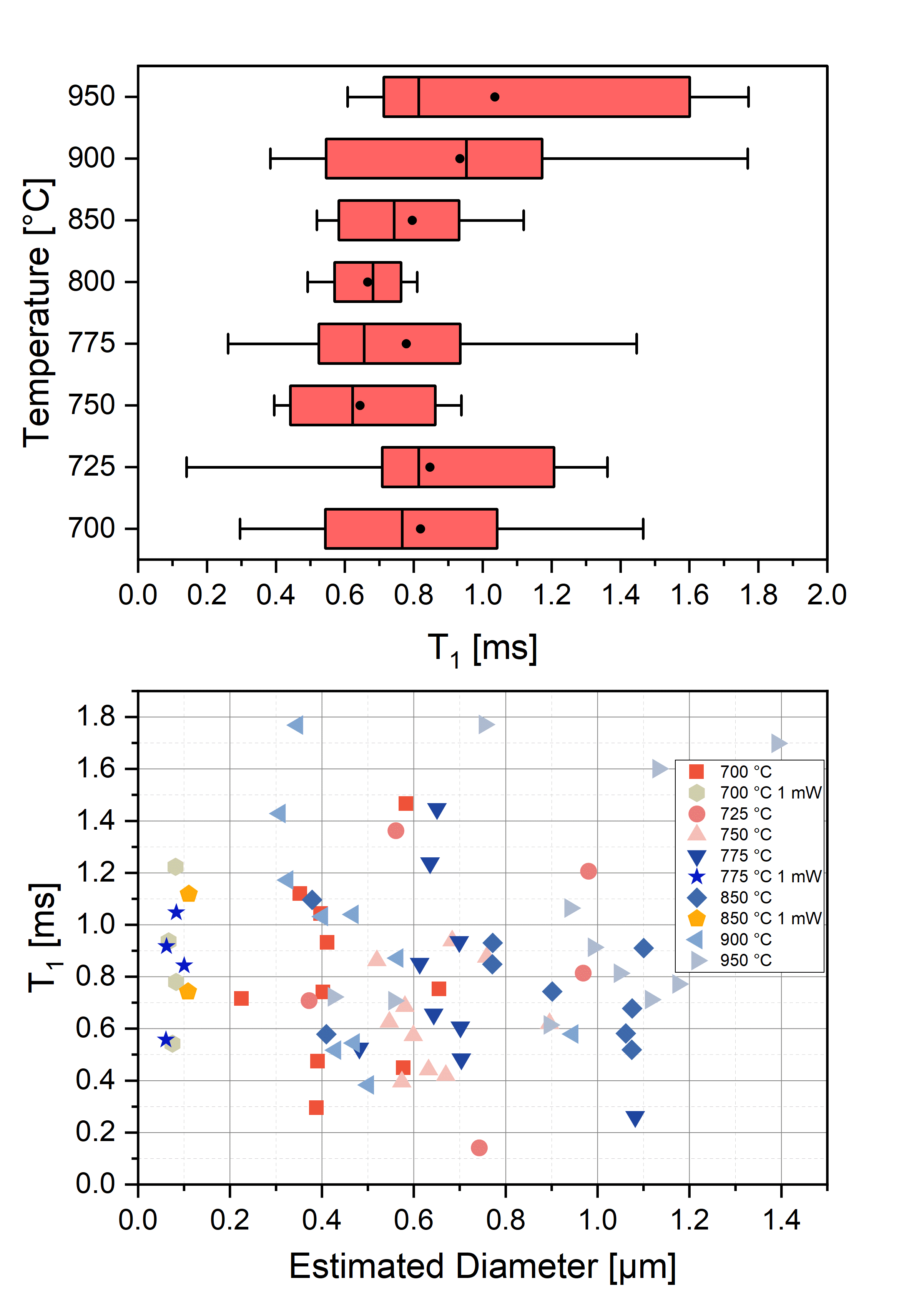}
    \captionof{figure}{Top: Overview of the range of measured\textit{T$_1$} times for all shell-doped samples, obtained with subtraction of a purely optical measurement with a microwave measurement. Mean relaxation times stay relatively consistent, Table S2. Bottom: Calculated particle size from its luminescence for each measured \textit{T$_1$} time. Unless specified, measurements were done at 50 $\mu$W. Two samples were measured at 1 mW as at lower power small particles blend into the background luminescence, see the Supporting Information. For measurements at 1 mW, specifically low-luminescent particles were chosen. While average \textit{T$_1$} is consistent between different samples, within the sample particles with strongly varying \textit{T$_1$} times can be found. This is most likely due to varying crystal morphology throughout the sample.}
    \label{fig:T1s}
\end{figure}

\textbf{Figure \ref{fig:T1s}} shows all \textit{T$_1$} measurements executed on the shell-doped nanodiamonds grown at different temperatures using a 50 µW laser. \textit{T$_1$} time ranges between 100 µs and 1800 µs. This spread could be a consequence of varying morphology between NDs, Figure \ref{fig:Morphology}, which depends on how a particle evolves from its nucleation to its final crystal shape, in particular depending on if \{100\} or \{111\} facets are more prominent, defects can be different. We achieve \textit{T$_1$} times on average of about 800 µs - a major improvement over HPHT nanodiamonds, which achieve \textit{T$_1$} times up 150 µs in particles of roughly 50 - 150 nm.\cite{Reineck} \textit{T$_2$} times do not show such improvement (see the Supporting Information), this we attribute to be due to the paramagnetic defect density,  such as P1 centres providing the spin bath. Indeed, according to the growth conditions, we expect the concentration of nitrogen to be around 10 ppm, while NV density is in the range of 100 ppb, whose ratio is significantly higher than for implanted and annealed HPHT diamond and is the limiting factor for \textit{T$_2$}.\cite{Wyk} We also show that in our samples, \textit{T$_1$} times seem relatively independent when increasing the mean particle size (Table S2). NDs smaller than 200 nm have to be measured with a higher laser power of 1 mW due to their weak luminescence.

\subsection*{Effects of Nitrogen Doping}
In this paragraph we compare the morphology of low-doped and shell-doped FNDs. A high nitrogen pulse is performed to engineer an outer, NV-doped shell, comparable to $\delta$-doping in bulk diamond. This technique is described in the Supporting Information. The effects of the high nitrogen doping on the FND morphology are studied using SEM and compared to the first set of FND samples grown under low N-doping conditions. Figure \ref{fig:Morphology} shows the difference in the surface morphology for low-doped as compared to shell-doped FNDs, grown at similar temperatures. The growth conditions for low doping were selected to lead to a mixture of \{111\} and \{100\} facets and cuboctahedron shapes at about 850 °C with an alpha parameter of roughly 1.75.\cite{WILD1994373} Whilst low N-doping leads to clean facets and nearly perfect crystal morphology, high N-doping across all growth temperatures causes major twinning. The thin diamond layer grown under the nitrogen pulse at the final step of the growth cycle shows re-nucleation and multi-twining growth on the \{111\} facets, while the \{100\} facets remain clean without re-nucleation. The effect of twinning in \{111\} facets is known, especially for high N$_2$ gas addition.\cite{KNUYT19981095} Due to this multi-twinning and re-nucleation, the high N-doped shell is highly defective, and we observe a decrease in relaxation time \textit{T$_1$} as compared to low-doped growth only, displayed in \textbf{Figure \ref{nitro}}. We could show that the growth at high nitrogen doping conditions leads to generation of surface defects, of which the integrated magnetic field reduces the \textit{T$_1$} time.\cite{Glenn2018} Also, the crystallographic defects enable other recombination channels and consequently the PL from the shell-doped layer is reduced. Another characteristic is that the \textit{T$_1$} measurements discussed in the next section are more difficult to execute due to the baseline drift. In particular, for example, by applying a MW $\pi$-pulse, the PL trace does not show the population inversion, as expected, shown in \textbf{Figure \ref{inv}}. However, by using the procedure (described in the following section), we can correct for the time-dependent baseline. The decrease in relaxation time is very likely to be therefore an effect induced by the increase in surface area due to multi-twining and due to an enhancement of surface spins density.\cite{KNUYT19981095} As mentioned in the previous section, in Figure \ref{fig:SEM-PL-SizeDistr} we have used the same fitting parameters to calculate the size distributions for the low-doped crystal growth as for the shell-doped size distributions. This makes us believe that the NV formation during the high nitrogen pulse, relative to the N concentration in the gas phase, is low and most of the nitrogen is incorporated as N$_s$.

\begin{figure}
    \centering
    \includegraphics[width=0.7\columnwidth]{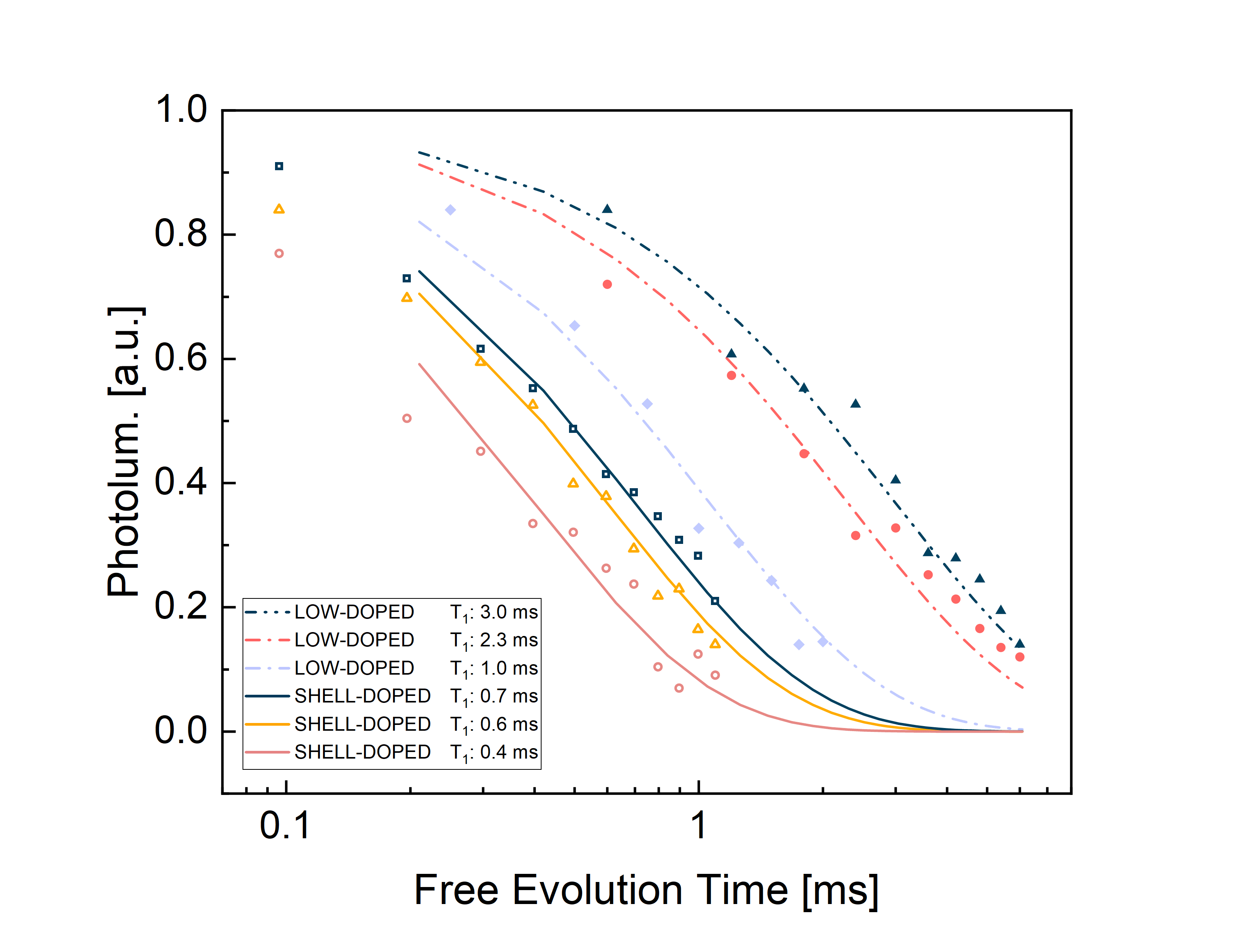}
    \captionof{figure}{Experimental data fits for shell-doped and low-doped nanodiamonds. Addition of a nitrogen pulse (shell-doping) to the growth sequence decreases the relaxation time from \textit{T$_1$} $\in [1 \text{ms},1.7 \text{ms}]$ to \textit{T$_1$} $\in[0.5 \text{ms},0.9 \text{ms}]$. Particles were calculated to be around 500 nm. Shell-doped nanodiamonds show lower \textit{T$_1$} times than the FNDs grown without shell-doping.}
    \label{nitro}
\end{figure}

\subsection*{Microwave \textit{T$_1$} measurements}
Nanodiamonds are vulnerable to NV charge state alternation.\cite{Barbosa} If charge state alterations are dynamic, they will introduce a baseline drift to the measurements, because the free evolution would, additionally to the spin state, also involve reaching charge state equilibrium. These alternations can occur via photoionisation of N, NV, and other defects present in FNDs, but also by redistributing the charge on the ND surface. This is quite natural, taking into account the possible band bending effects and their influence by surface termination.\cite{bandbend} As the band bending can range over several nm or more,\cite{Zhang2012} the charge alterations are naturally present. Band bending can occur for example due to adsorbates present at the surface.\cite{TACHIKI2000578} Here, we have measured on the naturally grown diamond surface without any additional treatment, as we preferred not to detach the FND from the silicon surface due to their small amounts and to prevent that any use of chemicals might lead to FND pollution by substrate surface etching. Diamond surface adsorbates can also change upon laser illumination, influencing the band bending,\cite{bandbend} for example, by flattening it by charge carrier generation. Also, PL from other defects can contribute to the detected signal. Therefore, during the measuring of \textit{T$_1$} relaxation time, the resulting PL curve contains both the nitrogen-vacancy relaxation and additional defects  relaxation. For these reasons, it is necessary to properly analyse the PL time dependence to distinguish between \textit{T$_1$} relaxation and parasitic effects.
\hfill \break

In order to exclude these parasitic effects, we perform measurements by using purely optical \textit{T$_1$} detection and also measurements with $\pi$-pulse microwave application.\cite{Schirhagl2014} Microwave radiation swaps the population of spin states 0 and 1, and this process is followed by a PL readout. If photoluminescence decay were a result of NV spin relaxation only, the relaxation curve would be mirrored after the MW-induced population swaps. Measurements of differences in time traces for purely optical and optical with MW driving have therefore been used \cite{PhysRevLett.108.197601} to fully decompose the PL decay spectra into spin flip dependent and independent processes. While for purely optical spin polarisation and decay measurements we anticipate a decaying PL curve to the steady state, when using $\pi$-pulse microwave driving in between the laser pulses we anticipate a rising PL with time.
\hfill \break

Our measurements have revealed the presence of an additional non-spin baseline relaxation, most likely originating at the charge alteration processes. Figure \ref{inv} shows the results of the three measurement sequences described in the Experimental Section, \textbf{Figure \ref{fig:Setup-Pulse}}, performed simultaneously. The “No MW” curve indicates the population of $m_s=0$ spin state after relaxation from $m_s=0$ state, the “MW After Free Evolution” curve indicates the population of $m_s=1$ state after relaxation from $m_s=0$, and the “MW Before Free Evolution” curve indicates the population of $m_s=0$ after relaxation from $m_s=1$. Subtraction of the curves that correspond to relaxation from $m_s=0$ should allow us to correct for the baseline. However, the difference between the curves changes very little, as they do not fully converge, causing a large error on the fitting parameters. Therefore, we choose to subtract the curve showing $m_s=1$ relaxation from the curve showing $m_s=0$ relaxation. Even though the relaxation path from $m_s=1$ is different than from $m_s=0$ state, and the decay from $m_s=1$ is biexponential, the population of $m_s=0$ state still changes with the same time constant.\cite{Myers} We solve rate equations to show that the \textit{T$_1$} is the same in both cases (see the Supporting information). The $T_1$ values presented in Figure \ref{fig:T1s} are calculated this way. We present a more complex background subtraction procedure in the Supporting Information.
\hfill \break

The fact that the two curves for relaxation from $m_s=0$ do not converge indicates that the state to which the system relaxes on the timescale of the measurement is not a full statistical mixture of the three triplet states. This is an unexpected result as thermal energy at room temperature should be higher than the splitting between the triplet states. We confirm that we have no leakage of laser light, as such leakage would indeed result in breaking the symmetry between the triplet states. We measure that AOM attenuation exceeds 50 dB for 10 µW laser power, which means that initialization time will increase from approximately 3 µs to more than 300 ms. Considering that the longest free evolution time in our measurements is 3 ms, the effects of laser leakage are negligible. Although the origin of this effect remains unconfirmed, our hypothesis is that it can be connected to, e.g., observed charge switching of NVs under illumination or in the dark,\cite{Bluvstein} or possibly to surface Fermi level pinning.\cite{bandbend} There can be an upwards band bending, as we observed, at the surface leading to higher population of NV$^0$ states.\cite{bandbend} When the band bending slowly relaxes (the charge redistribution at the surface is known to be slow, often in the range of minutes), the NV$^0$ will slowly transform to NV$^-$ state in which the $m_s=0$ population will be augmented \cite{ARTICLE:Wirtitsch2023}, the system will return to an equilibrium statistical mixture between the $m_s=0$ and $\pm$ 1 states, on the moment when the band bending relaxation stops. Notably, we do not observe this anomaly in nanodiamonds prepared with a standard method, i.e. HPHT nanodiamonds, \textbf{Figure} \ref{rd}. 

\begin{figure}[H]
    \centering
    \includegraphics[width=\columnwidth]{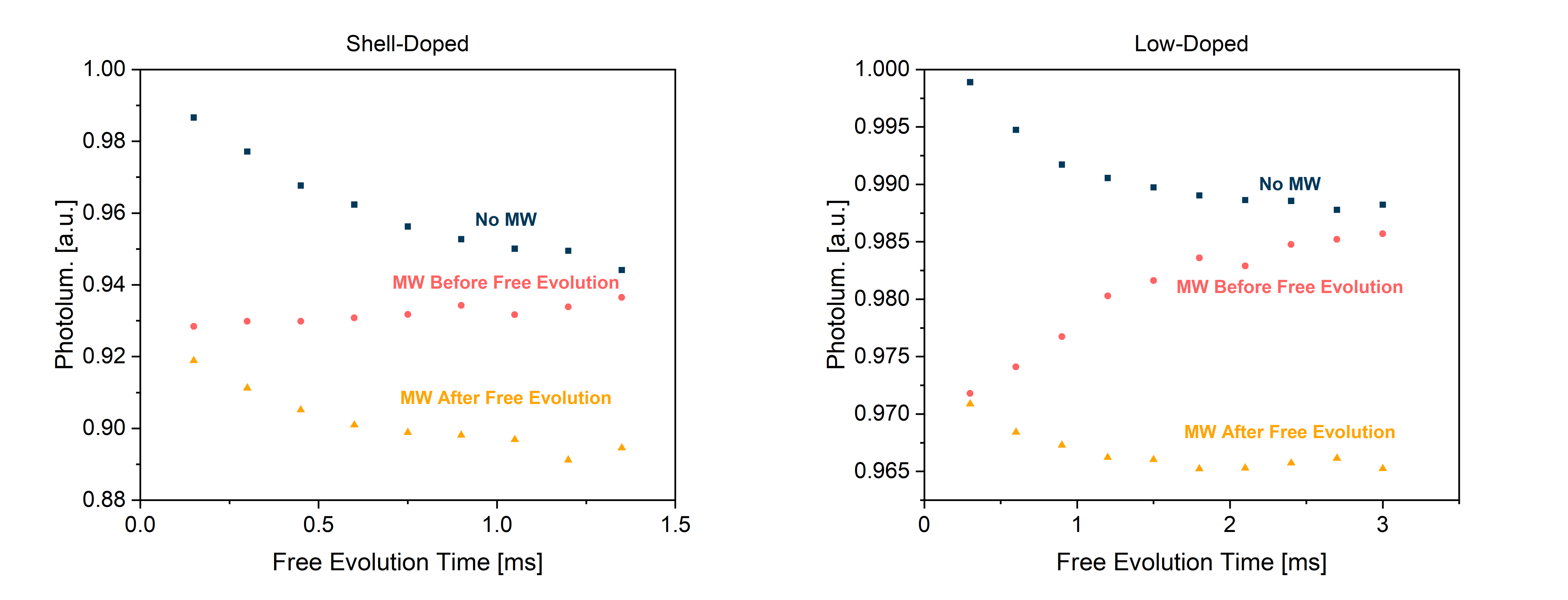}
    \captionof{figure}{Photoluminescence as a function of free evolution time for shell-doped (left) and low-doped (right) nanodiamonds. Low-doped FNDs show a stronger inversion of the relaxation-related decay.}
    \label{inv}
\end{figure}

\begin{figure}[h]
    \centering
    \includegraphics[width=0.5\columnwidth]{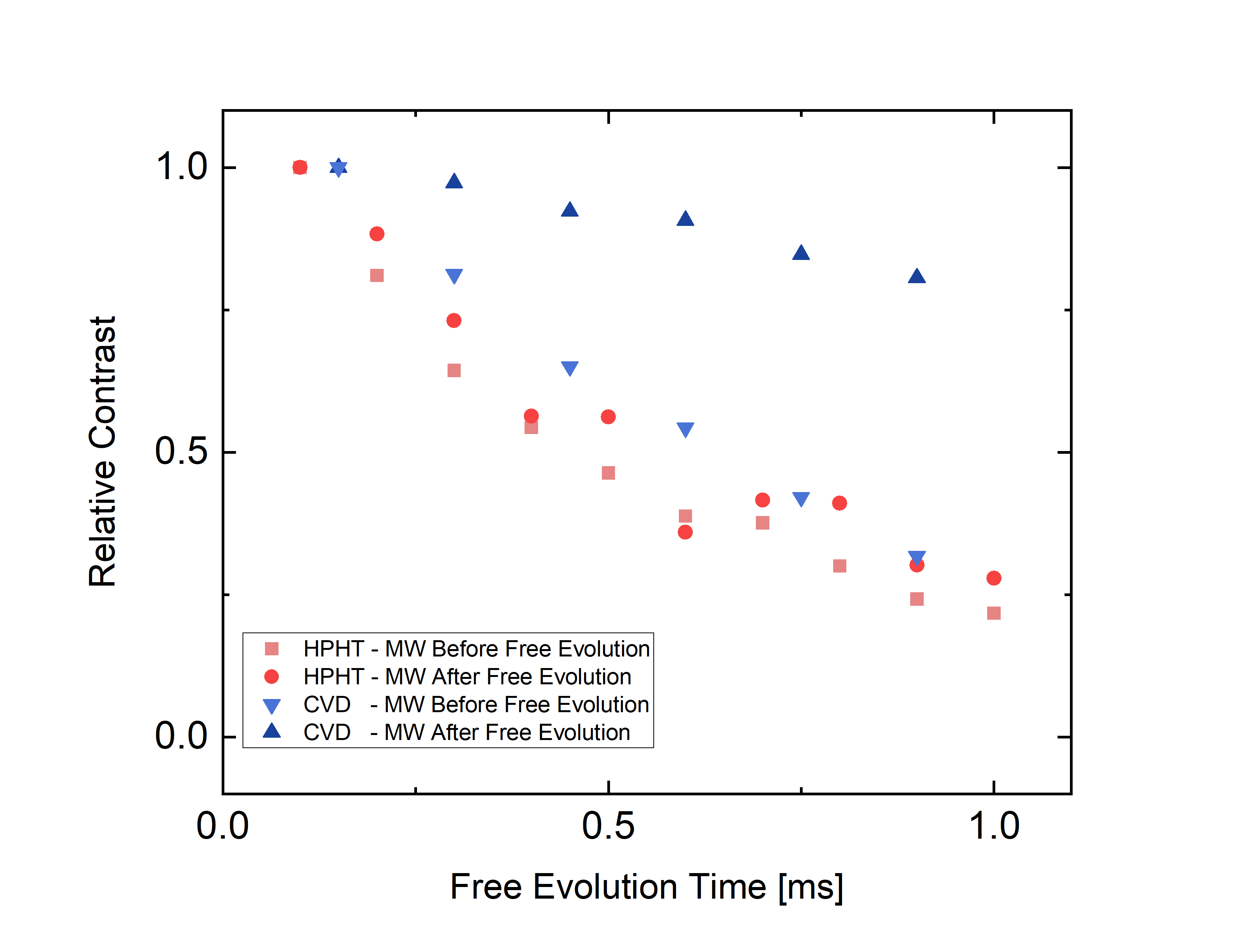}
    \caption{Difference between photoluminescence signal with and without microwave application. Data for the high \textit{T$_1$} sample diverges with respect to microwave pulse position, but data for the reference sample does not. For this comparison particles with similar \textit{T$_1$} are found in both samples.}
    \label{rd}
\end{figure}

\subsection*{Referencing}

For comparison of \textit{T$_1$} values in HPHT and CVD-FNDs, and also to verify the reliability of the \textit{T$_1$} protocols, a reference measurement is carried out on HPHT NDs with an expected \textit{T$_1$} time of roughly 100 µs.\cite{REDOXHPHT}\textbf{ Figure \ref{fig:HPHTNDS}}a shows \textit{T$_1$} measurements on 4 different nanodiamonds, three of which show relaxation times very close to the expected values. The dependence of these relaxation times on laser power is also measured, as shown in Figure \ref{fig:HPHTNDS}b. Measurements are performed on the same nanodiamond, but with varying laser power. Lower laser power shows a longer \textit{T$_1$} time (140 µs) compared to higher laser power (76 and 91 µs). However, it has to be noted that lower laser powers (if only shorter laser pulses are used) are less reliable, as full initialisation is not achieved, as seen in Figure \ref{fig:HPHTNDS}c. Additionally, the photoluminescence response of an NV centre depends on laser power, which is why in Figure \ref{fig:HPHTNDS}c, the readout of the thermal state (dashed lines) shows different pulse shapes for different laser powers. Although, the population inversion after a MW pulse properly occurs at higher laser power, showing that HPHT has a more stable surface. In that case, we used oxidised HPHT FNDs from \cite{STURSA}. On the other side, this highlights the need for using both the MW and optical \textit{T$_1$} measurements to correct for the FNDs that are not oxidised or in contact with a liquid or other environment.\\

\begin{figure}
    \includegraphics[width=\linewidth]{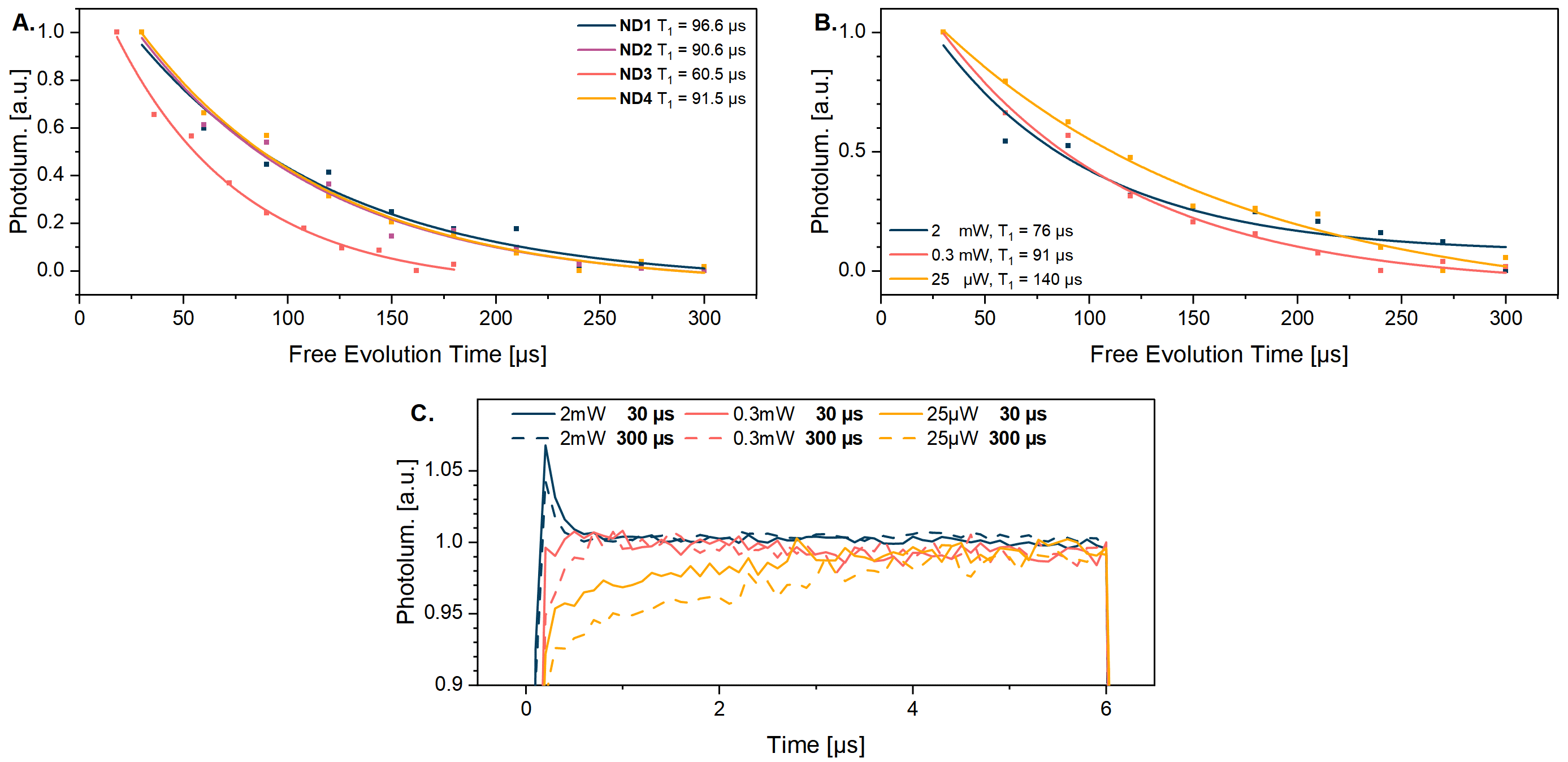}
    \captionof{figure}{A. Relaxation time measured for different HPHT nanodiamonds similar to those used in \cite{REDOXHPHT}. The difference between the nanodiamonds is within error, 100 µs is assumed to be the characteristic relaxation time for these samples. B. Relaxation time was measured for different laser powers. \textit{T$_1$} decreases with increasing laser power. At low powers, NV centre initialisation does not happen or happens to a small extent. C. Time-resolved photo-luminescence during readout pulse of 6 µs, before (solid) and after (dashed) full NV relaxation.}
    \label{fig:HPHTNDS}
\end{figure}

\section{Conclusion}

We fabricated fluorescent nanodiamonds with a high \textit{T$_1$} spin relaxation time, exceeding the current limits of HPHT nanodiamonds. To this end, we have performed CVD nanodiamond growth based on heterogeneous nucleation on pre-engineered surfaces. By measuring the FND size distributions and attributing them to their \textit{T$_1$} times, we show that we can achieve high mean \textit{T$_1$} times in particular 800 µs with a maximum over 1.8 ms for nanoparticle sizes averaging 60 nm at 700 °C.  This is about eight times higher than commonly found in commercially available HPHT nanodiamonds and already closer to the \textit{T$_1$} bulk diamond values. The regular diamond shapes and low density of defects are probably playing a major role in reaching high \textit{T$_1$} times.  We have found out that the \textit{T$_1$} time can easily be incorrectly evaluated from photoluminescence relaxation due to, probably, photoionisation of other defects or band bending changes occurring on the surface of NDs when illuminating with laser, as discussed in the Supporting Information. This is particularly relevant if FNDs are exposed to a time-varying environment, such as a biological environment. By measuring the SEM particle size and mapping the PL using developed dedicated software, we could determine the product of the N incorporation from the gas phase and the NV generation yield. The calculated N incorporation rate to FNDs is in the range of 1\%, suggesting that the NV incorporation still occurs mainly via \{111\} facets. Further on, we attempted to fabricate an NV-doped shell around the core ND by pulsing nitrogen at the final step of the growth; however, due to high doping, we created a highly re-nucleated \{111\} surface, which is defective. Thus, further optimisation to avoid the multi-twinning and re-nucleation for N-doped films together with a high incorporation is necessary to prepare smooth facets with high \textit{T$_1$} times with highly dense ensembles. Moreover, our particles exhibit approximately ten times lower photoluminescence compared to conventional HPHT nanodiamonds. Although previous studies have successfully conducted quantum measurements and tracking using nanodiamonds hosting single NVs within cells\cite{ARTICLE:McGuinness2011}, further improvement in the ND luminosity can be achieved by an additional irradiation treatment which would increase the NV concentration and reduce partially the N spin bath.\\
\hfill \break

While our nanodiamonds show enhanced relaxation times as compared to HPHT nanodiamonds, there is still further research needed to increase the nanodiamond yield. Our CVD method is a 2D method, so in order to achieve sufficiently large amount of FNDs for chemical processing one can use, for example, large area growth systems, such as Linear Antenna MW reactors. \cite{Taylor2011} FNDs can be grown on intermediate layers which are easy to chemically etch away to release FNDs and without the need to dissolve large amounts of the substrate material. We estimate that several milligrams of FNDs can be produced in one CVD deposition run. In addition one can grow on 3D structures such as porous media, nanotube arrays \cite{VLCKOVAZIVCOVA201861} or similar, to further increase the FNDs amounts deposited in one run.


\section{Experimental Section}
\textbf{MW Plasma Reactor.} A home-built plasma-enhanced chemical vapour deposition (PE-CVD) diamond reactor (See the Supporting Information), is used to grow nanodiamonds. The growth is controlled with custom-made Python software that communicates with the reactor's Programmable Logic Controller (PLC), which enables us to have precise control over growth parameters. The MW source consists of a 2.45 GHz microwave generator connected to a waveguide and has a maximum power output of 1.5 kW. Nanodiamonds were grown on Si substrates with sizes of 5 x 5 mm$^2$. The gas mixture and growth conditions are discussed below.\\

\textbf{Nanodiamond growth.}
ND nucleation pre-treatment on all samples is carried out after an initial ultrasonic cleaning using ethanol, followed by subsequent nano-roughening of the substrate with diamond powder using an ultrasonic vibration table (Vibromet Polisher) and 80-120 µm-sized diamond particles (Custodiam 80/120 Mesh) on Si substrates. The samples are again cleaned ultrasonically using ethanol and dried under nitrogen gas flow. \\
\indent Following their cleaning, the samples are placed within the diamond reactor, which is pumped down to its minimum base pressure of 1.1 E-8 mbar. Two sets of samples are grown: low nitrogen doped and FND equipped with a highly N-doped shell. The substrate temperatures are varied from 700 to 950 °C (see table S2 for the full list) as measured with a calibrated single wavelength pyrometer. The growth was carried out at a constant 1\% methane (4 sccm) addition in hydrogen (396 sccm), a pressure of 120 mbar, and a power of 1.0 kW for 10 minutes. For the growth of the low doped FND particles, we use H$_2$ gas with 10 ppm N$_2$. We also carried out our pulsed N-doping technique, where shell-doped samples have 5 sccm of N$_2$ added during the final 30 seconds leading to the growth of a $\delta$-doped shell with an intent to generate NVs close to the surface of the diamond particles. The full details of the pulsing technique can be found in the Supporting Information.\\

\textbf{Optical setup.} Figure \ref{fig:Setup-Pulse} depicts our quantum confocal setup used for measurements. It consists of a 532 nm laser beam modulated by an acousto-optical modulator (AOM) for the generation of laser pulses. A dichroic mirror is used to direct green light to the sample and red photoluminescence light from the sample to an avalanche photo-detector (APD) or a CCD spectrometer using a 0.95 N.A. objective. \textit{T$_1$} measurements are performed with a 650 nm long-pass filter, a 700 nm short-pass filter is used for imaging purposes to cut-out the SiV luminescence. The silicon substrate with grown nanodiamonds is glued to a PCB, over which a 50 µm thick wire is soldered. Microwaves can be applied through this wire, which is terminated with a 50 Ohm load.\cite{our} The setup is run with custom-made LabView software.

\begin{figure}
    \centering
    \includegraphics[width=0.5\columnwidth]{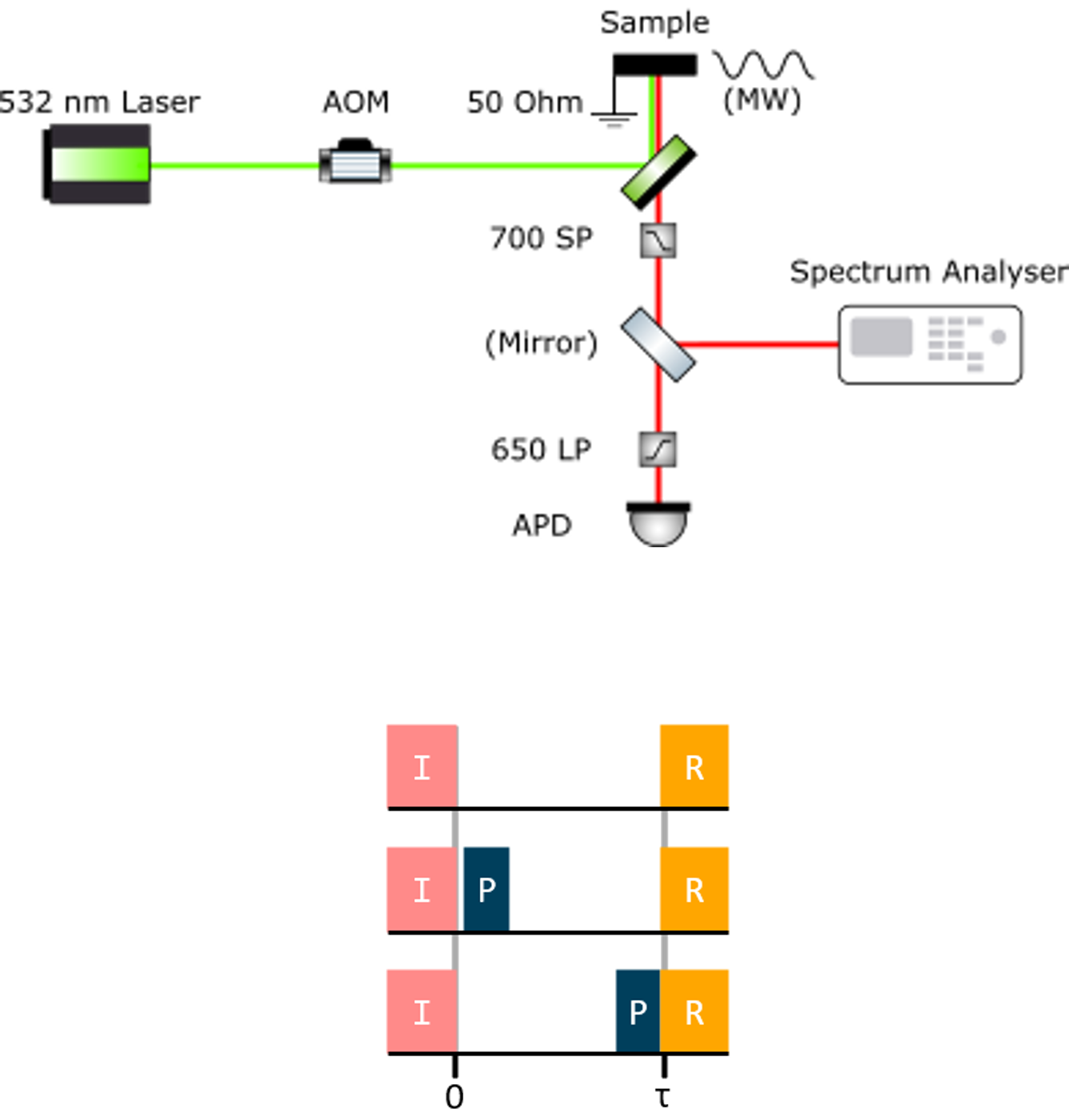}
    \captionof{figure}{Top: Schematic overview of the measurement setup. Standard confocal setup with the possibility to apply microwaves. Green laser (532 nm) goes through an acousto-optic modulator (AOM) to the sample, illumination from the sample is collected through wavelength filters to an avalanche photodiode (APD) or to a spectrometer. A dichroic mirror is used to combine sample irradiation with light collection. A microwave wire is placed over the sample surface for \textit{T$_1$} pulsed measurements. A 650 nm long pass and 700 nm short pass filters are used. No filter is used for measuring spectra. Bottom: Measurement sequences used for \textit{T$_1$} determination. Each sequence consists of an initialisation pulse (I) followed by free evolution time (\textit{$\tau$}), when no external perturbation is applied, and a readout pulse (R) in the end. The first sequence is a purely optical measurement, the other two sequences have an additional microwave $\pi$-pulse (P) right before or directly after free evolution.}
    \label{fig:Setup-Pulse}
\end{figure}

\textbf{High-resolution SEM.} High-resolution SEM imaging is performed with a Zeiss 450 FEGSEM with Gemini 2 Optics at 5 kV to study their morphology and to see the effect of nitrogen pulsing.\\
\textbf{Size distribution analysis.} Size distribution analysis is performed by several image analysis tools, in particular by combining high-contrast SEM images taken by a FEI Quanta and maximum-filtering photoluminescence imaging (see the Supporting Information).

\textbf{\textit{T$_1$} measurements.} To determine the relaxation times of the grown nanodiamonds, various optical and microwave pulsing schemes, as shown in Figure \ref{fig:Setup-Pulse}, are compared. A small external magnetic field is applied to split the resonances.

Initialisation of the NV centres in the nanodiamonds is achieved using a 532 nm laser pulse. After a time delay (\textit{$\tau$}), a readout of the colour centres is performed. During this delay, NV centres relax to their thermal equilibrium state with time constant \textit{T$_1$}. By varying \textit{$\tau$} we obtain a full relaxation curve for the specific nanodiamond. Optionally, a microwave pulse is  inserted either between the initialisation pulse and free evolution or between the free evolution and the readout pulse. The former changes the relaxation path from ms $\pm$ 1, while the latter merely swaps the state population before readout.\\
\indent Due to laser illumination and subsequent photoionisation, the charge state of the NV centre can change. As such, both optical and MW assisted \textit{T$_1$} measurements are performed, with further subtraction of the results in order to exclude the charge state alteration effects on the PL decay.\cite{Jarmol} Also, the baseline relaxation is important for some FNDs and we discuss this in detail in the Supporting Information.

\medskip
\textbf{Supporting Information} \par 
Supporting Information is available from the Wiley Online Library or from the author.

\medskip
\textbf{Acknowledgements} \par 
The author would like to acknowledge the following projects: Diamond for chip-based quantum detection in bioelectrode multi-electrode recordings of human iPSC-derived neurons and axonal networks (R-11434), i-BOF; Project Number G0A0520N from FWO; EU Quantera Project Maestro; EU Project Amadeus, Advancing the market uptake of diamond defects quantum sensors, Grant agreement ID: 101080136\\

M.G. acknowledges project No. 101038045 (ChemiQS): This project has received funding from the European Union’s Horizon 2020 research and innovation programme.\\

Jeroen Prooth and Michael Petrov contributed equally to this work.

\textbf{Conflict of Interest} \par 
We declare no conflict of interest.

\newpage
\medskip

%
\bibliographystyle{MSP}
\bibliography{papers.bib}

\newpage







\begin{figure}
\textbf{Table of Contents}\\
\medskip
  \includegraphics[width=5.5cm]{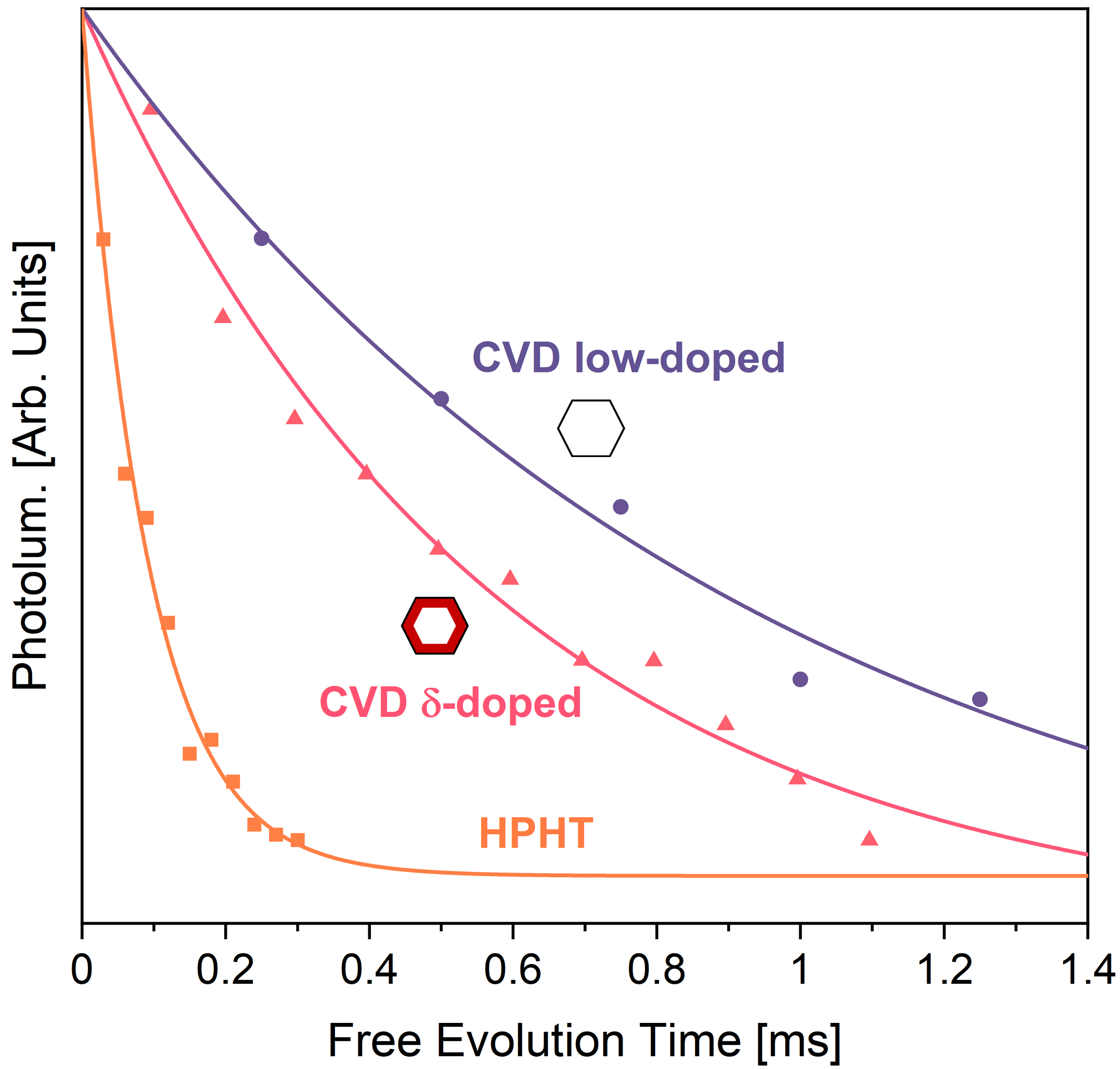}
  \medskip
  \caption*{On average, an eight-fold increase in the spin relaxation time is achieved for fluorescent nanodiamonds grown by CVD as compared to commercial HPHT nanodiamonds. The techqniue is based on heterogeneous diamond nucleation. Also we were able to create a thin NV $\delta$-doped diamond shell around nanodiamond particle, aimed at sensing applications.}
\end{figure}

\end{document}


\section*{Supporting Information}

\section{Nitrogen Pulsing}
\begin{figure}[H]
    \centering
    \includegraphics[width=0.75\columnwidth]{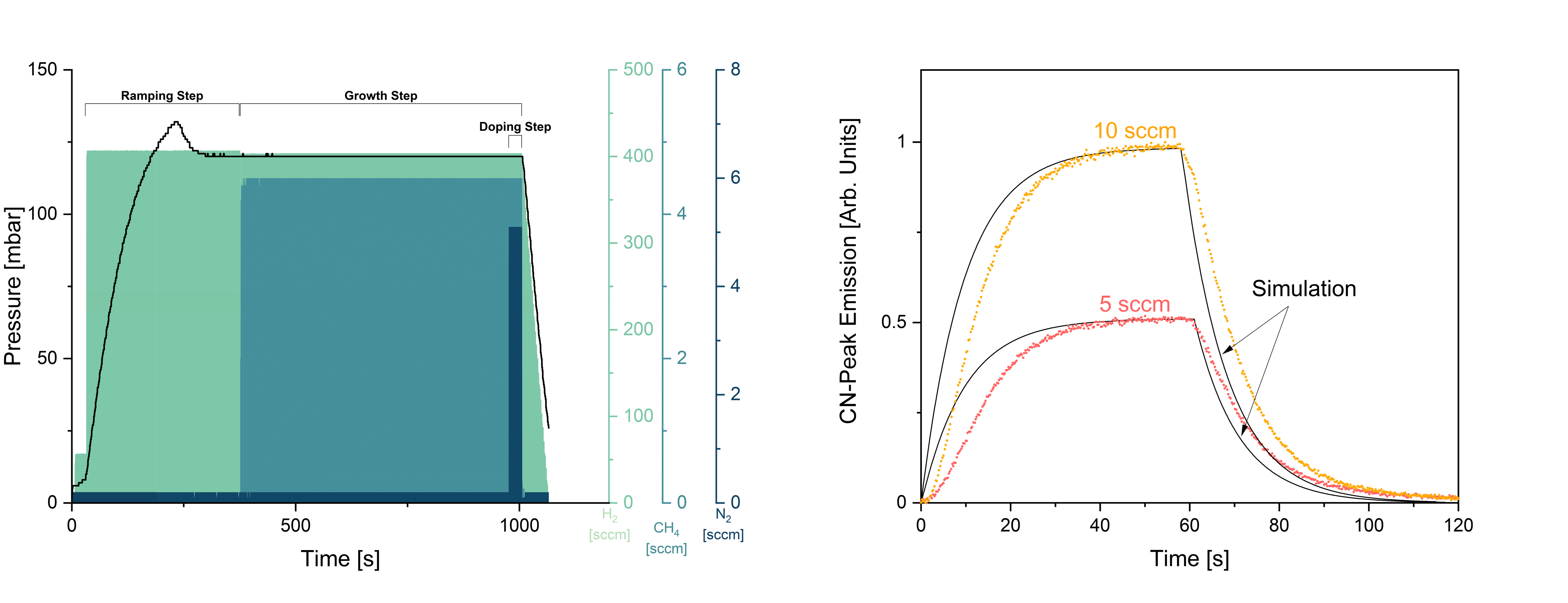}
    \captionof{figure}{\textbf{Left}: Typical growth profile for a shell-doped sample. The "Doping" step is omitted for low-doped samples. The full growth duration stays the same in both cases. \textbf{Right}: Nitrogen concentration profile as measured by optical emission spectroscopy of the CN line and corresponding simulations for 5 and 10 sccm respectively. It takes several seconds for the nitrogen concentration to reach saturation within the chamber.}
    \label{fig:SI-GrowthPulsed}
    \end{figure}

For the growth of $\delta$-doped shells around the bulk nanodiamond, a nitrogen pulsing scheme is introduced, figure \ref{fig:SI-GrowthPulsed}. Here the full growth profile is shown, and is the same for all shell-doped samples. For the low-doped samples, the "Doping" step is omitted, but the duration is kept the same. Power is a constant 1 kW, and pressure 120 mbar. The total flow into the chamber is kept at 400 sccm (H$_2$ + CH$_4$ + N$_2$ = 400 sccm). During the "Ramping" step, no growth is achieved, as no methane is introduced into the chamber. During this step, the reactor is allowed to reach steady-state. After, in the "Growth" step, methane is introduced at 4 sccm and diamond is allowed to nucleate and grow. The final step, the "Doping" step, introduces nitrogen at a constant rate of 5 sccm for 30 seconds. Notice that there is a background signal of N$_2$ flow, however the valve to the source gas is closed outside of the "Doping" step.

\begin{figure}[H]
    \centering
    \includegraphics[width=0.75\columnwidth]{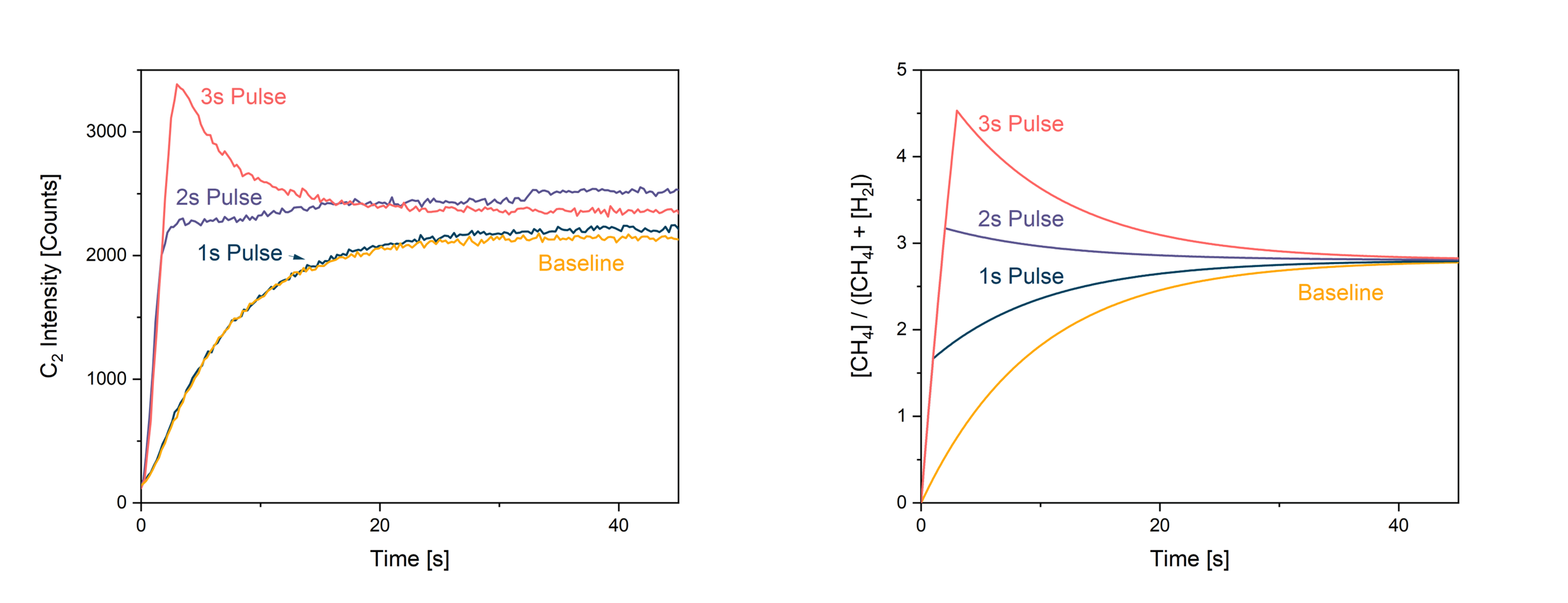}
    \captionof{figure}{\textbf{Left}: Measurements done at the C$_2$ emission line for the corresponding pulse duration within the plasma. The steady-state time of methane concentration can be reduced to only two seconds compared to the baseline of roughly 30 seconds.  \textbf{Right}: Simulations of pulsing CH$_4$ for 0, 1, 2 and 3s respectively at 72 sccm and then reducing to 12 with a total H$_2$ + CH$_4$ flow of 400 sccm. The difference in 1s Pulse is most likely due to the error in the flow controller response time.}
    \label{fig:SI-CH4Pulse}
    \end{figure}

While nitrogen is introduced at 5 sccm, this does not result in an immediate saturation of 1.25\% within the chamber. This effect is shown by measuring the optical emission spectrum of the CN line at 386.9 nm, figure \ref{fig:SI-GrowthPulsed}. Assuming that the CN line corresponds to the amount of nitrogen within the plasma, we show that it takes several seconds before the final concentration is reached. More complex pulsing schemes using multiple steps can be introduced, for nitrogen as well as methane. These pulsing schemes allow for much faster saturation times within the plasma, for example, pulsing methane initially at much higher flows for a few seconds, figure \ref{fig:SI-CH4Pulse}, reduces the time needed to reach state state from 30s to only 2s within the plasma.

\section{Additional Scanning Electron Microscopy of sub-100 nm Particles}
\begin{figure}[H]
\centering
\includegraphics[width=0.9\columnwidth]{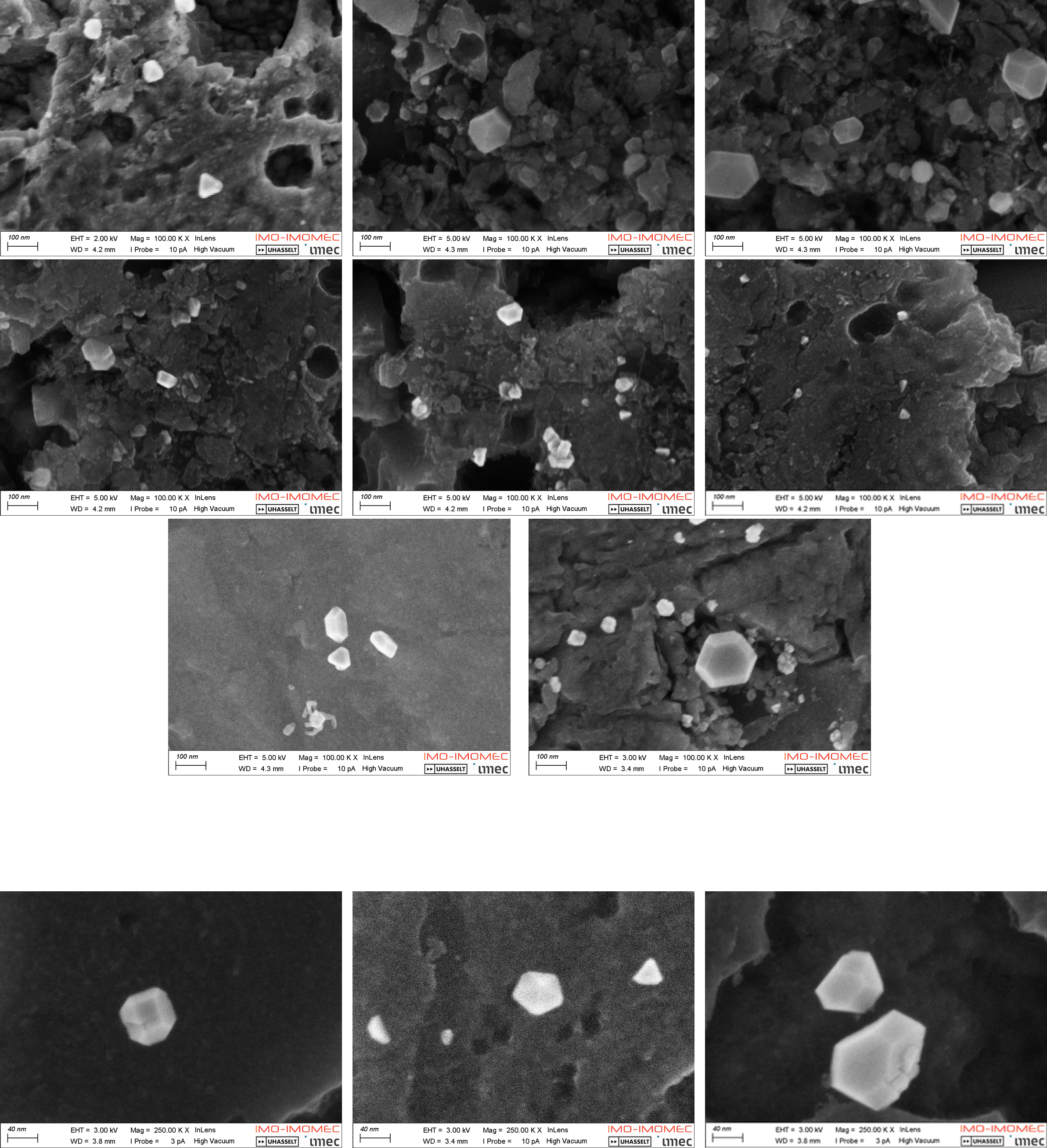}
\captionof{figure}{SEM images of additional low-doped nanodiamond particles, showing that even sub-50 nm particles show good morphology. Slight charging takes place, limiting the resolution of the SEM which can achieve single-digit nm resolution, figure S4. Taken with a Zeiss 450 FEGSEM with Gemini 2 Optics.}
\end{figure}

\begin{figure}[H]
\centering
\includegraphics[width=0.9\columnwidth]{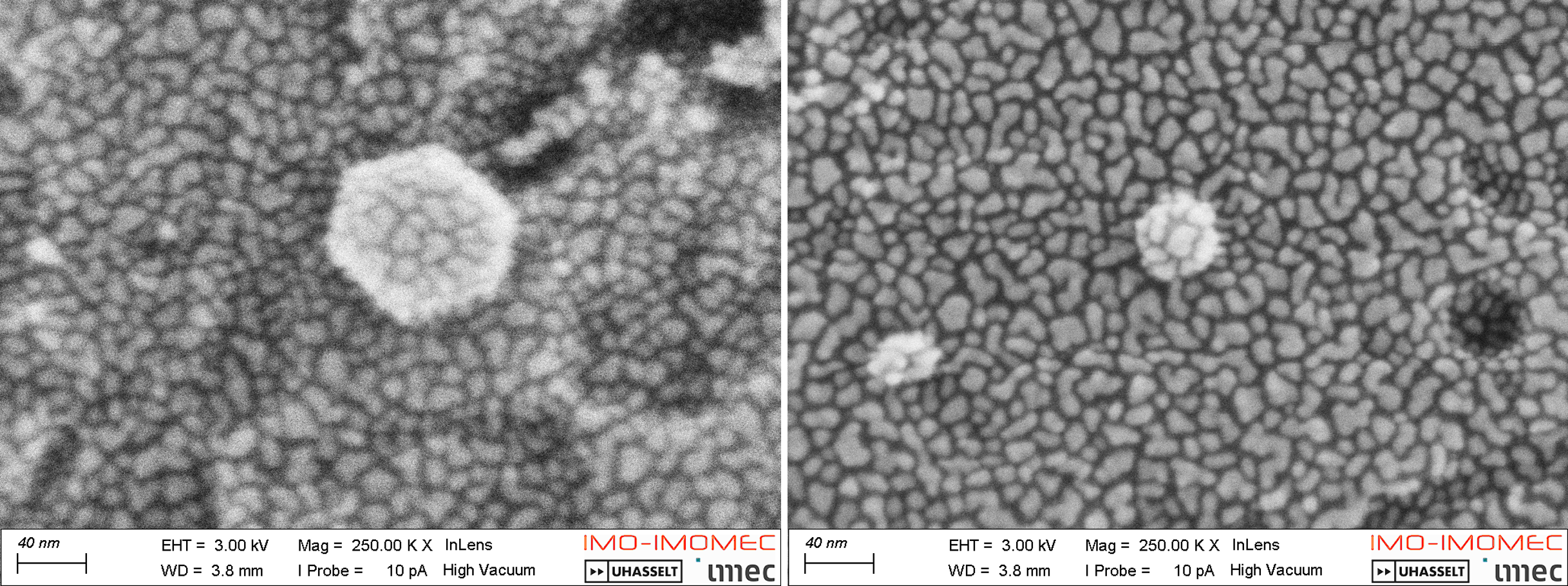}
\captionof{figure}{Au-Pd (2 nm) was sputtered on top of the substrate to combat charging effects. Au-Pd layer beads up instead of forming a thin film, however, image does show capabilities of SEM, with about 2.5 nm resolution in between beads.}
\end{figure}

\section{Image Analysis}

\subsection{Scanning Electron Microscopy Size Distributions}
\label{subsection:SEM_Ana}
Image analysis is done using a combination of the open-source tools ImageJ and Python 3. Scanning Electron Microscope (SEM) images are taken using a FEI Quanta electron microscope. The workflow is outlined below for a shell-doped sample grown at 950 °C. Several images are taken of each sample under high contrast and at a constant magnification, resulting in a corresponding pixel size of roughly 8x8 nm$^2$, limited by the chosen magnification. The images are then analysed using the following ImageJ functions, in order:

\begin{description}
    \item[Gaussian Blurring] A Gaussian Blur is applied to the image to filter out shot-noise. Drawback is that it smooths particle edges, slightly increasing the error on particle size. Sigma = [0..., 3] for our particles.
    \begin{figure}[H]
    \centering
    \includegraphics[width=0.75\columnwidth]{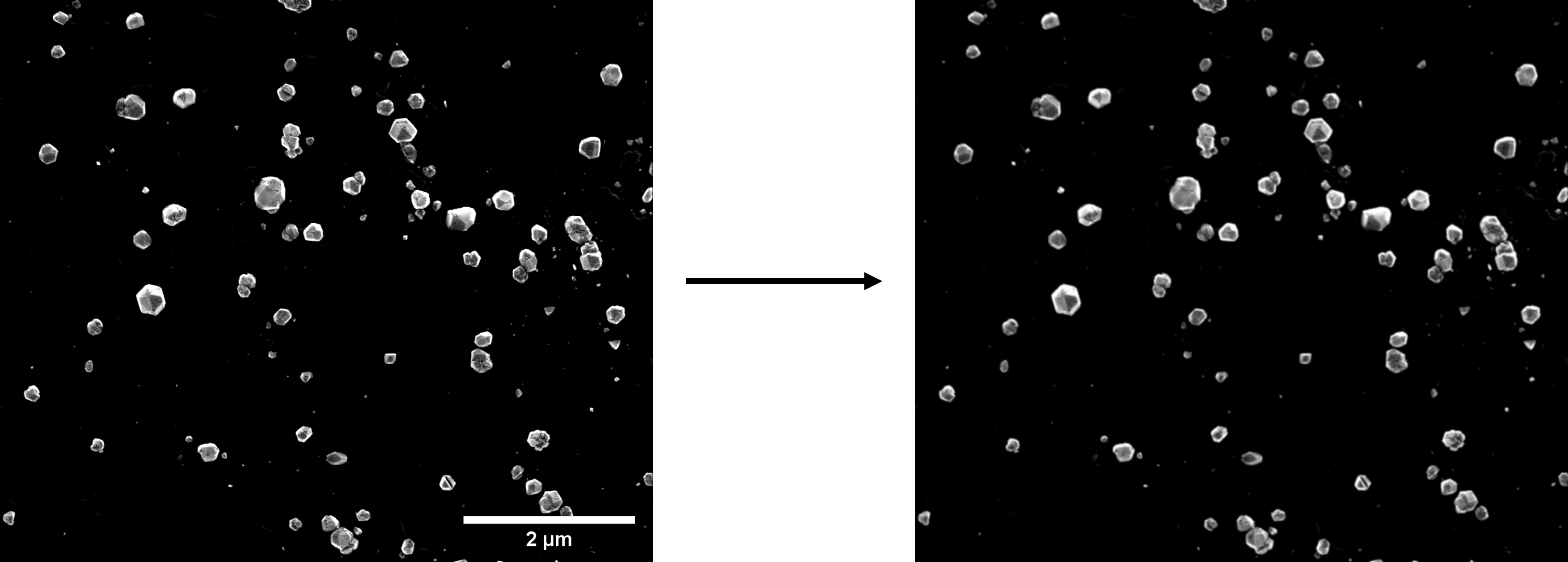}
    \captionof{figure}{Blurring of sigma = 1 applied to an SEM image. Edges of particles are smoothed and shot-noise is mostly removed.}
    \end{figure}
    
    \item[Thresholding] Further processing requires a binary image, which is achieved by thresholding. In our case, the "Default Dark", "Huang", or "Intermodes" algorithm is chosen depending on which achieves the cleanest results for the current set of images analysed. The result is a black-and-white image where the particles are black, and the background is white. 
    \begin{figure}[H]
    \centering
    \includegraphics[width=0.75\columnwidth]{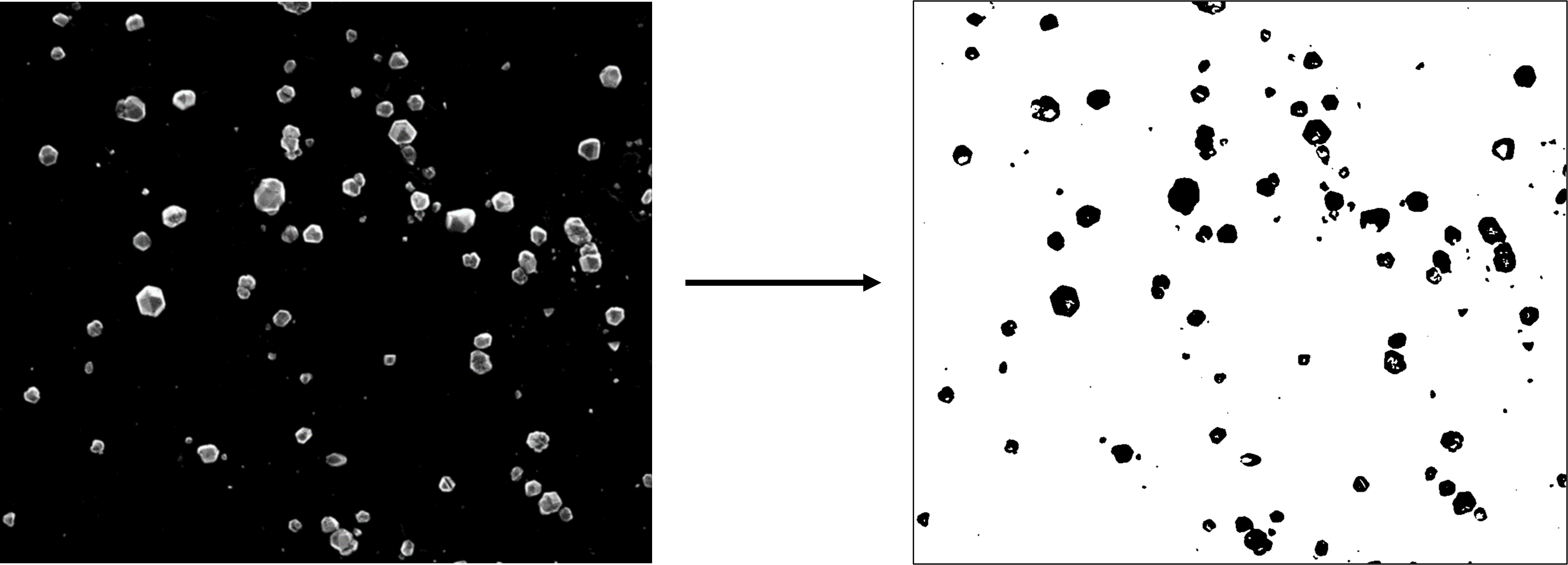}
    \captionof{figure}{Intermodes thresholding algorithm applied resulting in a binary image with particles in black and background in white.}
    \end{figure}
    
    \item[Fill Holes] Basically self-explanatory. Fills small holes left behind due to thresholding in the now black particles. Not a magic catch-all. Prevents that some particles are not detected.
    \begin{figure}[H]
    \centering
    \includegraphics[width=0.75\columnwidth]{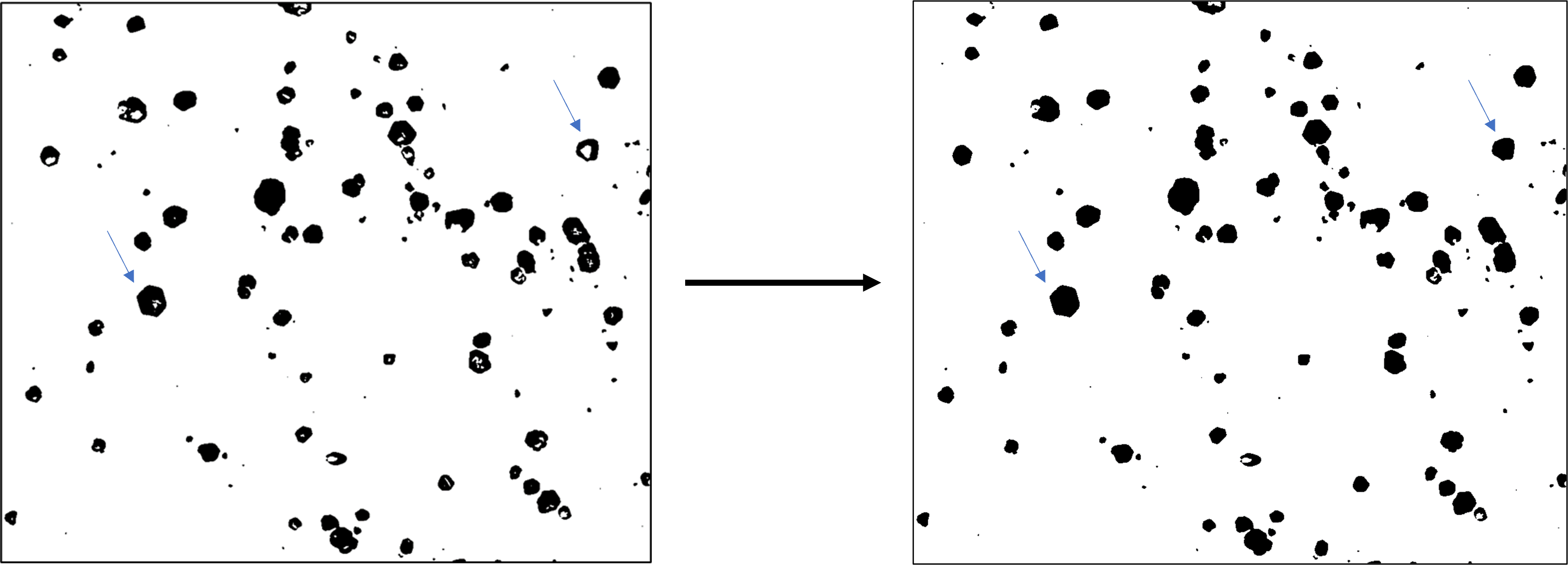}
    \captionof{figure}{"Hollow" particles are filled. They need to have an already closed boundary for this to work.}
    \end{figure}
    
    \item[Watershed] Incredibly important step. When particles are close to each other, the thresholding could result in one larger black particle as they are merged. Watershedding is a very powerful technique which tries to separate them back into their (almost) original boundaries. Not a magic catch-all.
    \begin{figure}[H]
    \centering
    \includegraphics[width=0.75\columnwidth]{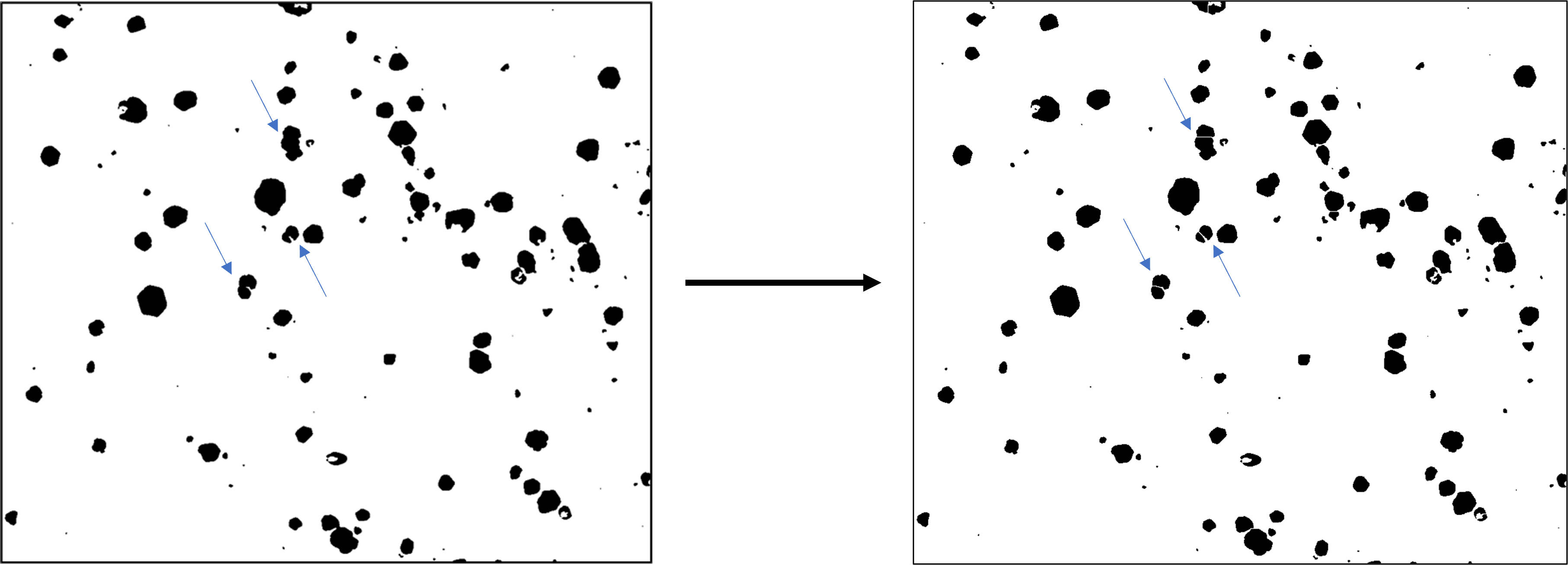}
    \captionof{figure}{Watershedding tries to break apart connected particles into their individual constituents.}
    \end{figure}
    
    \item[Analysis] ImageJ has an easy-to-use 'Analyse Particles' function. It requires a well-defined binary image and basically calculates the area of each particle. To actually filter the particles, one can ask ImageJ to exclude those below a minimal pixel-size (noise or false-positives) and requiring a certain circularity ('moon-shaped' particles are excluded). We chose a minimum pixel-size of 5 and a minimum circularity of 0.4 to 0.6. As ImageJ is also given the size of each pixel, we know roughly the cross-section area of each particle. For the final calculation of the diameter, we assume the particles are fully circular. 
    \begin{figure}[H]
    \centering
    \includegraphics[width=0.75\columnwidth]{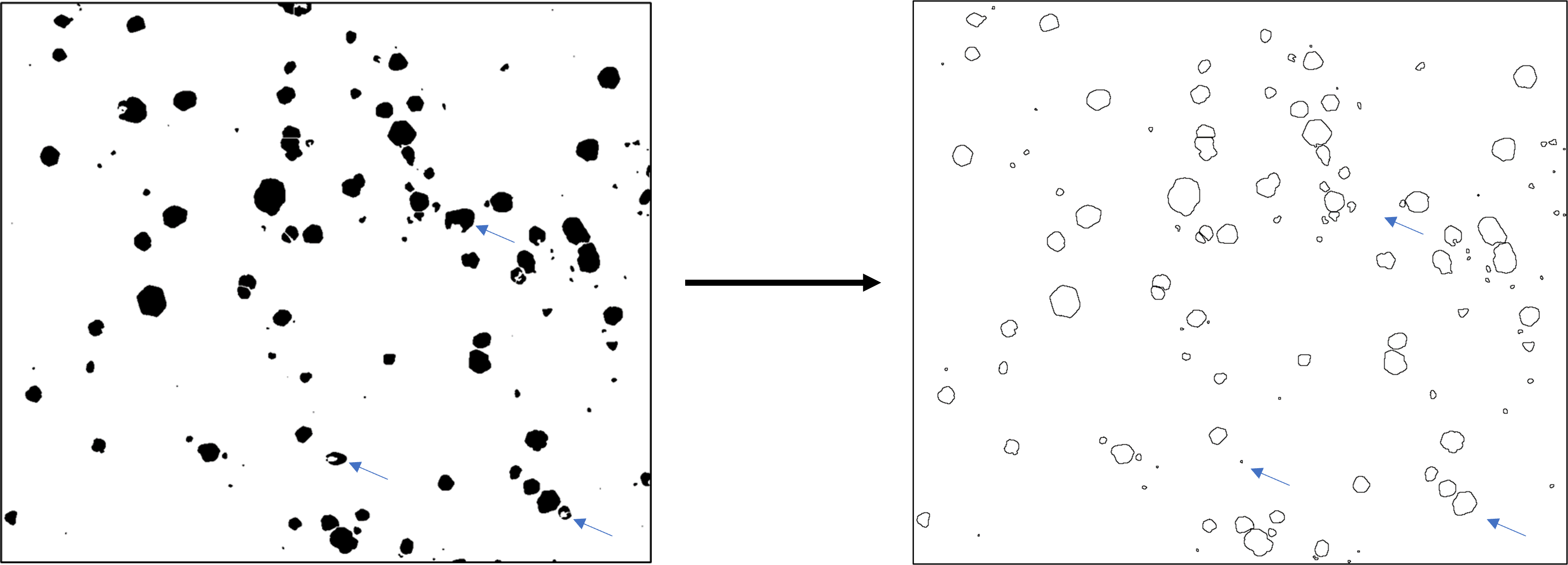}
    \captionof{figure}{Final ImageJ step where the binary image is analysed. The output is a table (here visually represented by the outlines) of particles with their area already calculated by ImageJ. Further processing happens with Python. Notice that false-positives do happen (left arrow), however these are minimised so do not influence the end result significantly. Weirdly-shaped, or "moon-shaped", particles are removed.}
    \end{figure}
\end{description} 

This processing is done for several (up to 8) images for each substrate automatically with an ImageJ script. Resulting images are then visually checked by overlaying them with their corresponding SEM image. If necessary, parameters such as the thresholding algorithm, the minimal pixel size, and circularity can be modified until satisfactory. The full analysis of the up to 8 SEM images is very quick, and usually takes only a minute in total. The end result of all analysed images are then used to create the size distribution. 

\begin{figure}[H]
\centering
\includegraphics[width=0.75\columnwidth]{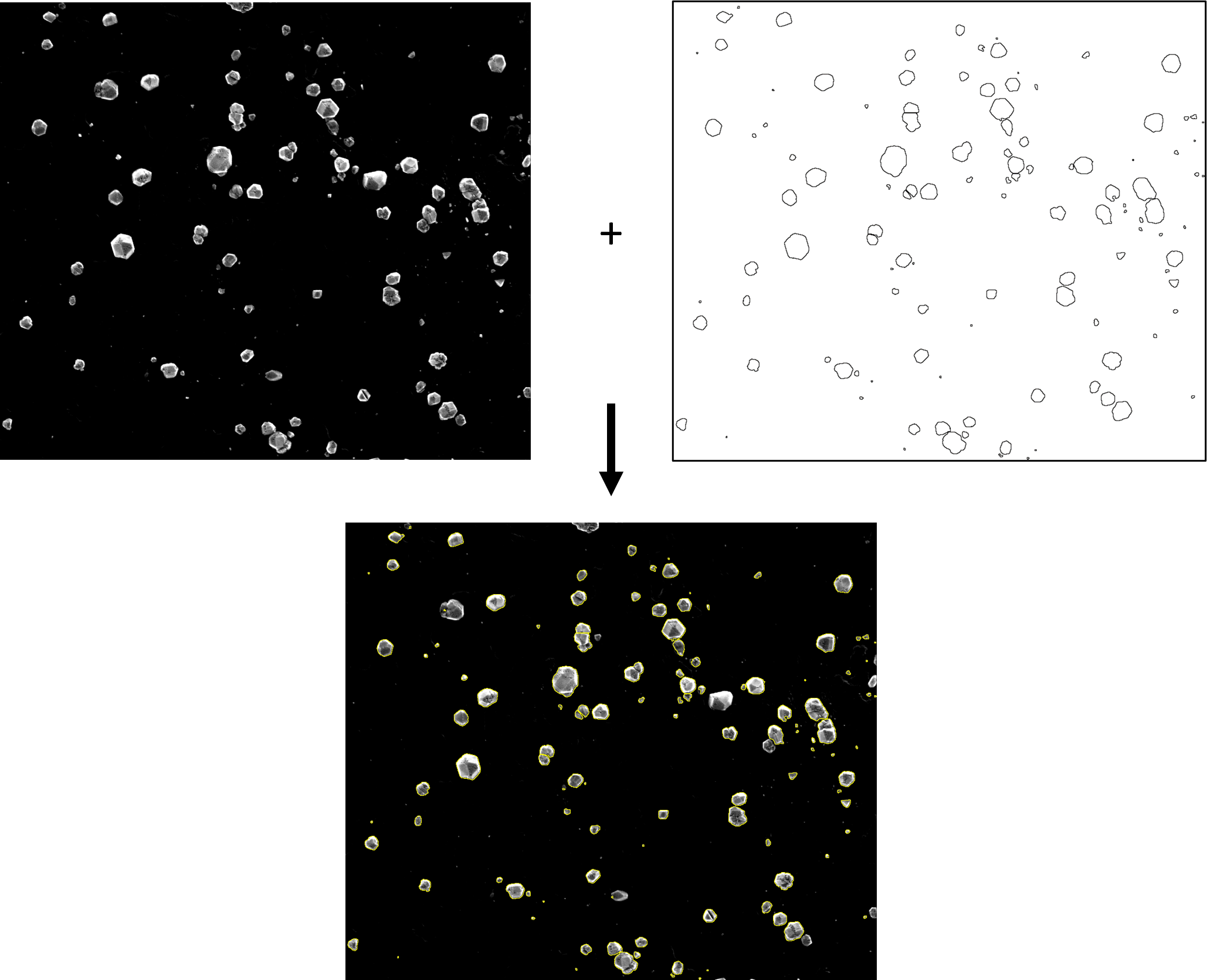}
\captionof{figure}{Visual check is done by overlaying the corresponding original and analysed ImageJ images. If needed, analysis parameters can quickly be changed and the full process re-done in only a minute.}
\end{figure}

\begin{figure}[H]
\centering
\includegraphics[width=0.75\columnwidth]{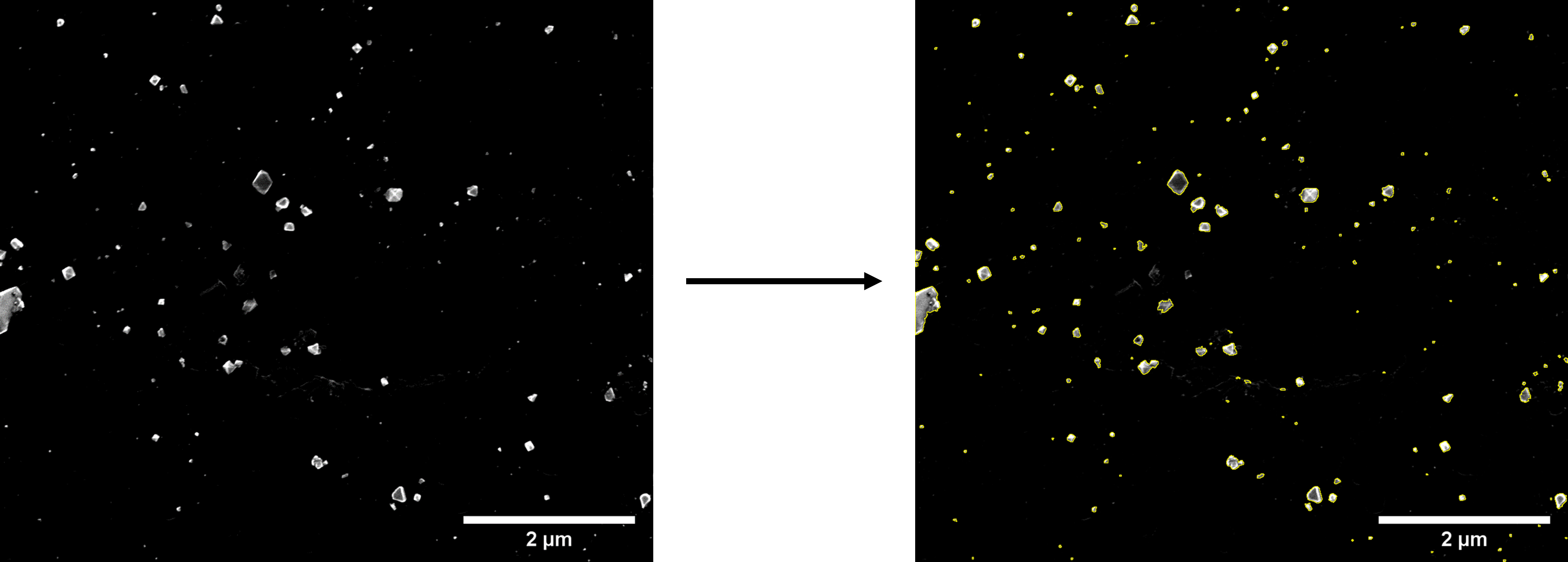}
\captionof{figure}{A similar analysis done for a shell-doped sample grown at 700 °C. It shows that particles grown at 700 °C are much smaller in size.}
\end{figure}
\subsection{Photoluminescence Distributions}

As in subsection \ref{subsection:SEM_Ana}, the photoluminescence (PL) distributions (and subsequent PL size distributions) are achieved using Python 3. For the PL distributions, the SciPy library is used.
Of each substrate, six 10 x 10 µm$^2$ scans at different positions are done under 50 µW of laser power. It is assumed that a particle in the PL map is defined by a local maximum value. As our particles are the same order of magnitude as the beam diameter of our laser, roughly 250 nm, if the full particle is illuminated, then a maximum amount of PL counts is obtained. Partial illumination results in lower counts. Thus, six PL maps are analysed as follows, in order:

\begin{description}
     \item[Gaussian Blurring] A Gaussian Blur is applied to the image to filter out shot-noise. Sigma = 1. Similar to Gaussian blurring in section \ref{subsection:SEM_Ana}, however here it slightly decreased the maximum value as it will be averaged with its nearest neighbours. 
    \item[Maximum Filter]A Maximum Filter is applied to the PL data. This filter defines the maximum of a local region, which is visually represented below. This local region is defined by a threshold and size (neighbouring). We have used a threshold of 3E3, slightly higher than background counts, and local area size of 5 neighbours during our analysis. 
    \begin{figure}[H]
    \centering
    \includegraphics[width=0.75\columnwidth]{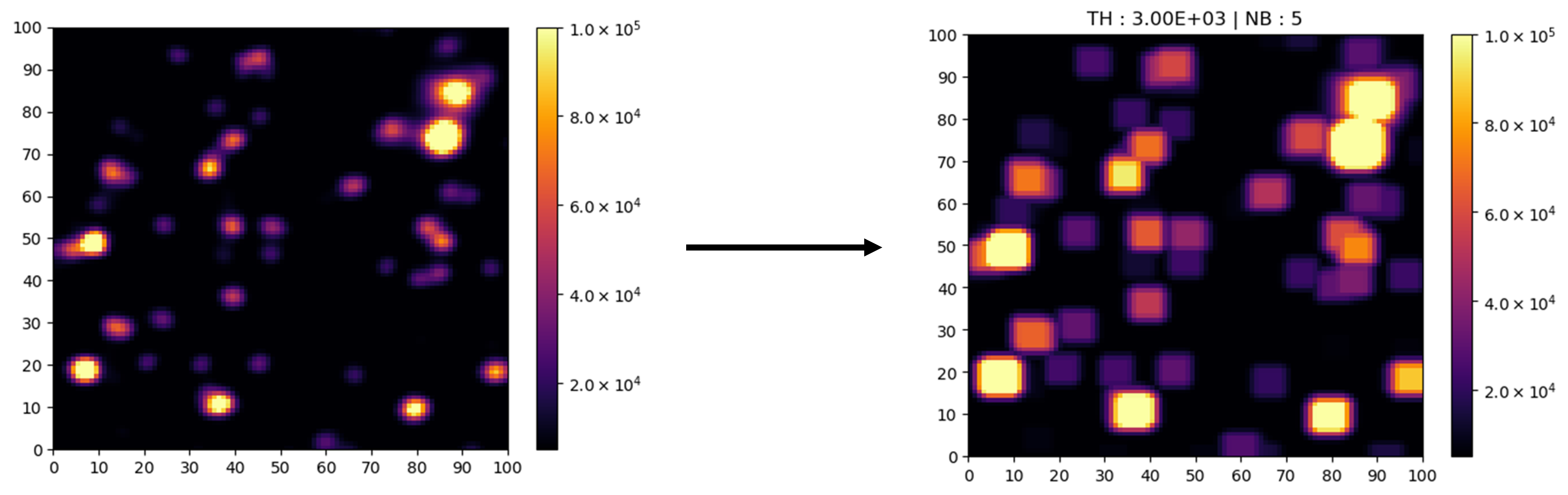}
    \captionof{figure}{A maximum filter is applied to a PL map, with a threshold of 3E3 counts (anything below is ignored) and assuming 5 neighbours. The amount of neighbours defines the size of the area of the local maximum. The maps are 100x100 pixels.}
    \end{figure}
    
    \item[Overlaying] Combining both previous images, it is possible to find the actual maximum value of that region simply by comparing both data sets. The result is a list of values describing each point where a local maxima was found. This is done six times for each sample, and the results for each are combined to get a PL distribution.
    \begin{figure}[H]
    \centering
    \includegraphics[width=0.375\columnwidth]{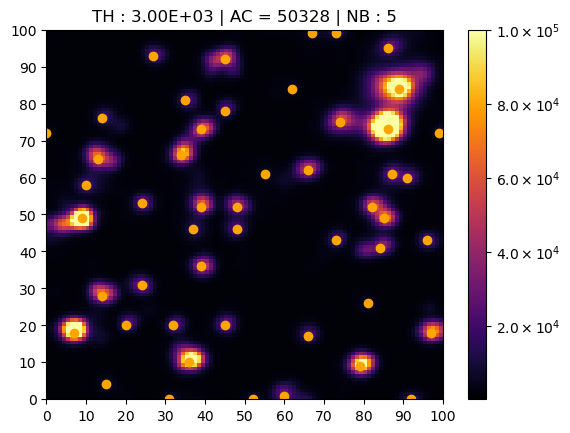}
    \captionof{figure}{Resulting maxima found by the algorithm are represented by an orange dot.}
    \end{figure}
    
\end{description}

\subsection{Combining SEM and PL Distributions}

To calculate the size distribution from the PL distribution, we assume that the nitrogen is uniformly incorporated into the nanodiamond and the amount of NVs thus corresponds with the size or volume of the particle. This means that we can use the size distribution as analysed by SEM to fit the PL distribution. \\
The diamond unit cell contains a total of 8 carbon atoms and is a face-centered cubic cell with side \textbf{a} = 3.58 \si{\angstrom}. We neglect bond lengthening from N incorporation. For a constant laser power, we assume that the photoluminescence, \textbf{I$_s$}, of a single NV is also constant (e.g. independent on growth conditions). Similarly, the total nitrogen content as compared to methane, \textbf{$\frac{[N_2]}{[CH_4]}$}, is also constant. Only the nitrogen incorporation rate \textbf{N$_i$} and the nitrogen to NV conversion rate \textbf{N$_c$} are temperature dependent, and thus different for different growth conditions. The total photoluminescence, \textbf{I$_{tot}$}, of a particle then simply depends on the amount of NV centres present, or:

\begin{equation}
    I_{tot} = nI_s
\end{equation}
with n the amount of NV centres. To find n, we need to find the number of carbon atoms replaced by nitrogen and subsequently converted to NV centers. Assuming the measured PL volume is spherical, the amount of carbon atoms \textbf{C} within a particle of diameter \textbf{d} is:

\begin{equation}
    C = m^3  \frac{atoms}{m^3} = \frac{4\pi(\frac{d}{2})^3}{3} \frac{8}{a^3} = \frac{4\pi d^3}{3a^3}
\end{equation}

 The total rate of carbon atoms replaced by nitrogen atoms and converted to NV centers, \textbf{R}, is equal to the amount of N$_2$ compared to the amount of CH$_4$ present during growth multiplied by the rate at which these nitrogen atoms are incorporated into the diamond lattice multiplied by the rate at which they are converted to NV centers:

\begin{equation}
    R = \frac{[N_2]}{[CH_4]}N_cN_i
\end{equation}

\noindent Combining previous three equations, we find:

\begin{equation}
    I_{tot} = CRI_s = \frac{4\pi d^3 \frac{[N_2]}{[CH_4]}N_cN_i}{3a^3}I_s
\end{equation}
\noindent Solving for the diameter of the particle we find:
\begin{equation}
\centering
d = \sqrt[3]{\frac{3I_{tot}a^3}{4\pi \frac{[N_2]}{[CH_4]} I_{s} N_{i} N_{c}}}
\label{SI:PLFormula}
\end{equation}
with \textbf{$\frac{[N_2]}{[CH_4]}$} and \textbf{I$_s$} having the same values for each substrate, as these are independent on growth conditions. Applying this formula to the PL distribution, we simply need to fit for N$_i$ x N$_c$ assuming that the peaks within the SEM distribution correspond to the peaks in the converted PL size distribution.

\begin{figure}[H]
\centering
\includegraphics[width=0.5\columnwidth]{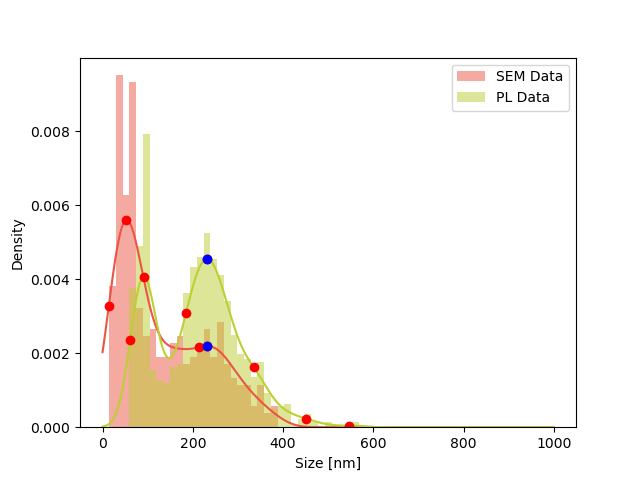}
\captionof{figure}{Size distribution of the SEM and PL data. Lines are the probability density function for the corresponding histogram. Dots signify the shoulders and peaks, with blue dots the points chosen so that the PL data can be fitted to the SEM data.}
\label{SI:HistFitting}
\end{figure}

Fitting is performed with a custom Python3 script. It looks for peaks and shoulders on the probability density function of the corresponding histogram. We can then fit for N$_i$ x N$_c$ by choosing which peak or shoulder we think should visually correspond to the peak on the other distribution. Figure \ref{SI:HistFitting} gives a visual representation for a sample grown at 775 °C. Red and blue dots represent the peaks and shoulders found on the distributions, with the blue dot signifying the peak in each distribution that should correspond to the other. An overview of the found N$_i$ x N$_c$ values can be found in table \ref{SI:PLFitParams}.

\sisetup{output-exponent-marker=\ensuremath{\mathrm{e}}}
\begin{table}[H] 

\setlength\tabcolsep{0pt} 
\begin{tabular*}{\textwidth}{l @{\extracolsep{\fill}}
                            *{3}{S[table-format=1.4]}} 
\toprule
& \textbf{Fitted on low-doped sample set} & \textbf{Fitted on shell-doped sample set} \\
{\textbf{Temp. [°C]}} & {N$_i$ x N$_c$} & {N$_i$ x N$_c$} \\ 
\midrule
\textbf{700 °C} &  & \num{2.21E-4}   \\ 
\textbf{725 °C} & & \num{1.22E-4} \\ 
\textbf{750 °C} & \num{2.1E-4} & \num{1.49E-4} \\ 
\textbf{775 °C} &  & \num{7.63E-5}  \\ 
\textbf{800 °C} & \num{6.29E-5} & \num{8.96E-5} \\ 
\textbf{850 °C} & \num{3.29E-5} & \num{3.14E-5}  \\ 
\textbf{900 °C} & \num{5.44E-5} & \num{7.34E-5}  \\  
\textbf{950 °C} & & \num{7.39E-5}  \\ 
\bottomrule
\end{tabular*} 
\caption{Overview of the N$_i$xN$_c$ values found by fitting the PL distributions to the SEM distributions of low- and shell-doped data sets. I$_s$ was taken to be 1300 and $\frac{[N_2]}{[CH_4]}$ to be 0.001.}
\label{SI:PLFitParams} 
\end{table}

\begin{figure}[H]
\centering
\includegraphics[width=0.5\columnwidth]{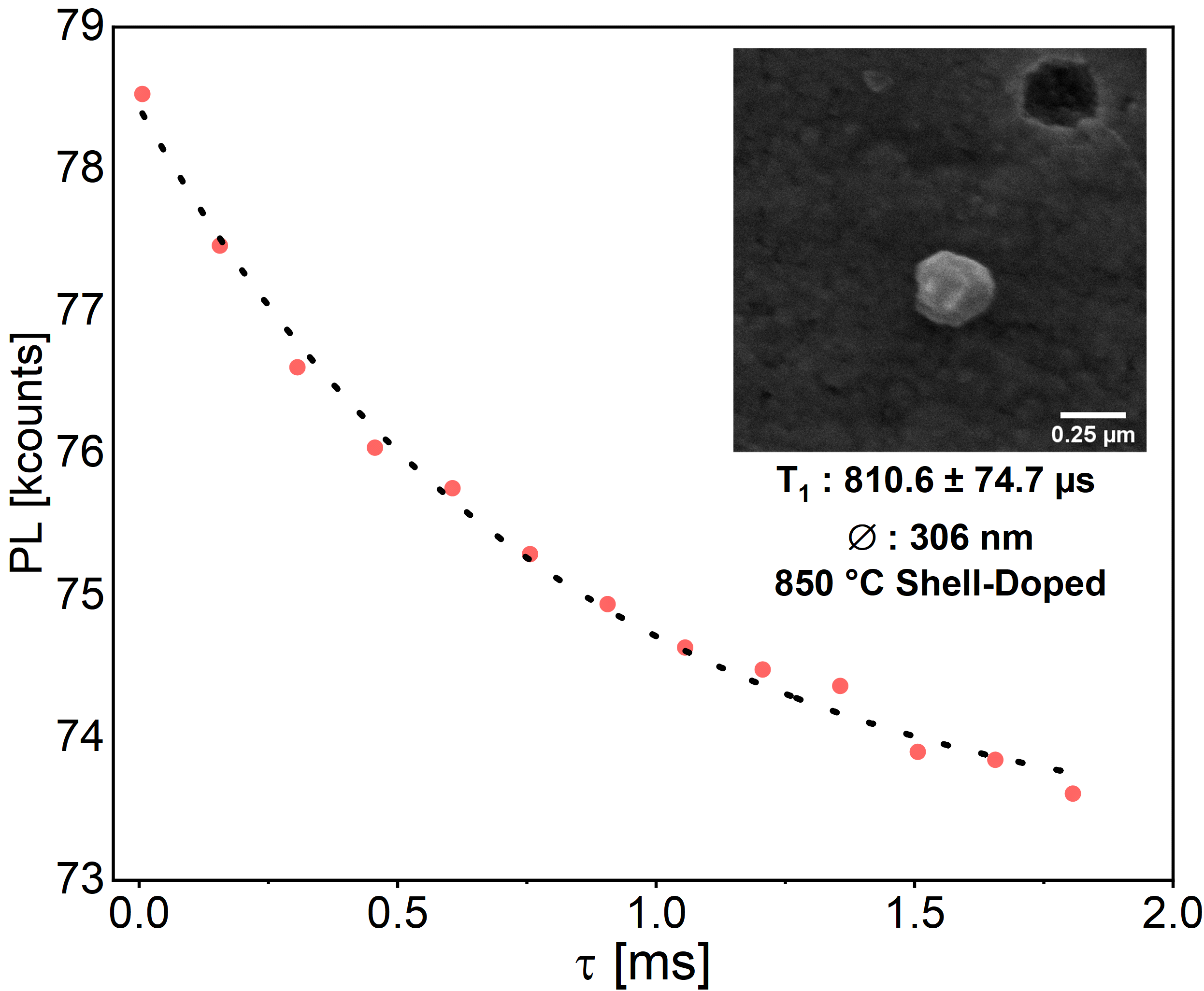}
\captionof{figure}{A solitary shell-doped nanodiamond grown at 850 °C with a measured \textit{T$_1$} time of 810.6 µs and diameter of 306 nm. The measured photoluminescence was 78 kcounts at 50 µW. A corresponding diameter of roughly 280 nm was found using equation \ref{SI:PLFormula} and the fitting parameters in table \ref{SI:PLFitParams} for 850 °C.}
\label{SI:MeasuredParticle}
\end{figure}

To confirm that the obtained parameters in table \ref{SI:PLFitParams} correspond to reality, we applied them to a known shell-doped particle, figure \ref{SI:MeasuredParticle}. Using formula \ref{SI:PLFormula} and the values for a sample grown at 850 °C found in table \ref{SI:PLFitParams}, we find a calculated diameter of roughly 280 nm, closely matching the measured SEM diameter of 306 nm.

\subsection{Statistical Data Particles}

\begin{table}[ht]
\begin{adjustbox}{width=0.5\columnwidth,center}
    \centering
    \begin{tblr}{colspec = {c c c c}, row{1} = {tableheader!75}}
         Temp. [°C] & Mean [nm] & 75\% Max. [nm] & 99\% Max. [nm] & MGR [nm min$^{-1}$] &  \textit{T$_1$} [µs]\\
         \textbf{700} & 58  & 71 & 255 & 5.5 & 799\\
         \textbf{725} & 123  & 185 & 447 & 11.7 & 847\\
         \textbf{750} & 133 & 192 & 390 & 12.7 & 644\\ 
         \textbf{775} & 134 & 217 & 373 & 12.8 & 778\\ 
         \textbf{800} & 141 & 195 & 351 & 13.4 & 666 \\  
         \textbf{850} & 193 & 290 & 434 & 18.4 & 765\\ 
         \textbf{900} & 159 & 244& 394 & 15.1 & 934\\ 
         \textbf{950} & 140 & 205 & 421 & 13.3 & 1035\\ 
    \end{tblr}
\end{adjustbox}
    \caption{Overview of the statistical data of the size distributions of the shell-doped nanodiamonds as measured by SEM and their mean \textit{T$_1$} time. A small trend toward higher mean and max sizes when growing at higher temperatures is noted, signifying an increase in growth speed, as can also be seen by the mean growth rate (MGR), \textit{T$_1$} times stay relatively consistent, trending slightly higher at 900 and 950 °C. The 75\% and 99\% max signify the percentage of nanodiamonds smaller than the indicated size.}
    \label{tab:sizedistr_table}
\end{table}

\subsection{Growth on Different Substrates}
\begin{figure}[H]
\centering
\includegraphics[width=0.5\columnwidth]{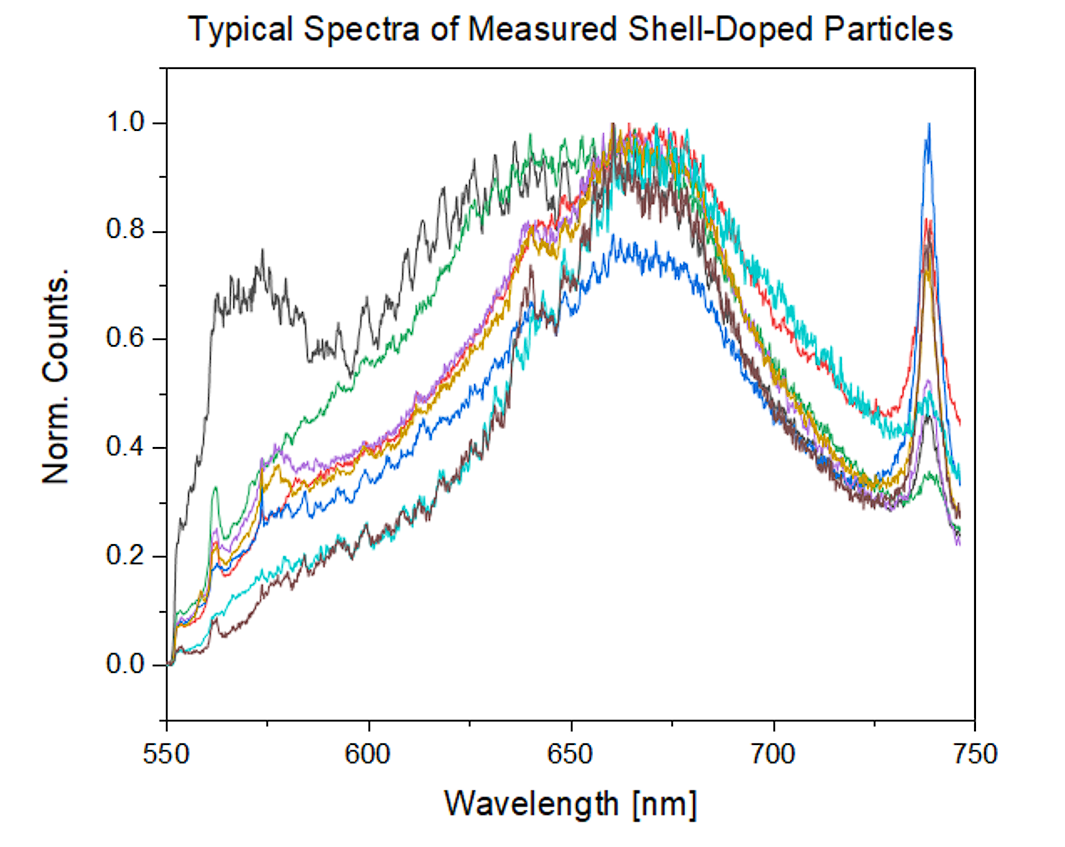}
\captionof{figure}{Spectra of several measured NDs grown on a silicon substrate.}
\label{SI:SiliconVacancies}
\end{figure}

\begin{figure}[H]
\centering
\includegraphics[width=0.5\columnwidth]{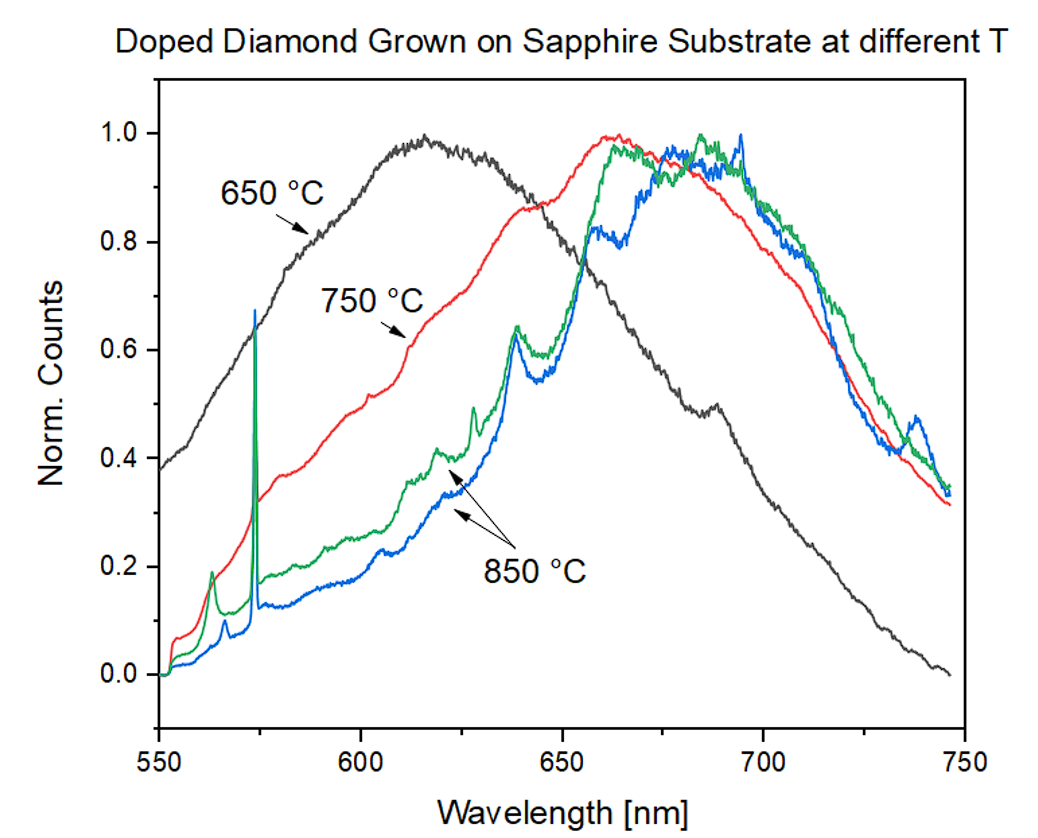}
\captionof{figure}{Spectra of several diamond crystals grown on a sapphire substrate.}
\label{SI:SapphireGrowth}
\end{figure}

\begin{figure}[H]
    \centering
    \includegraphics[width=0.5\columnwidth]{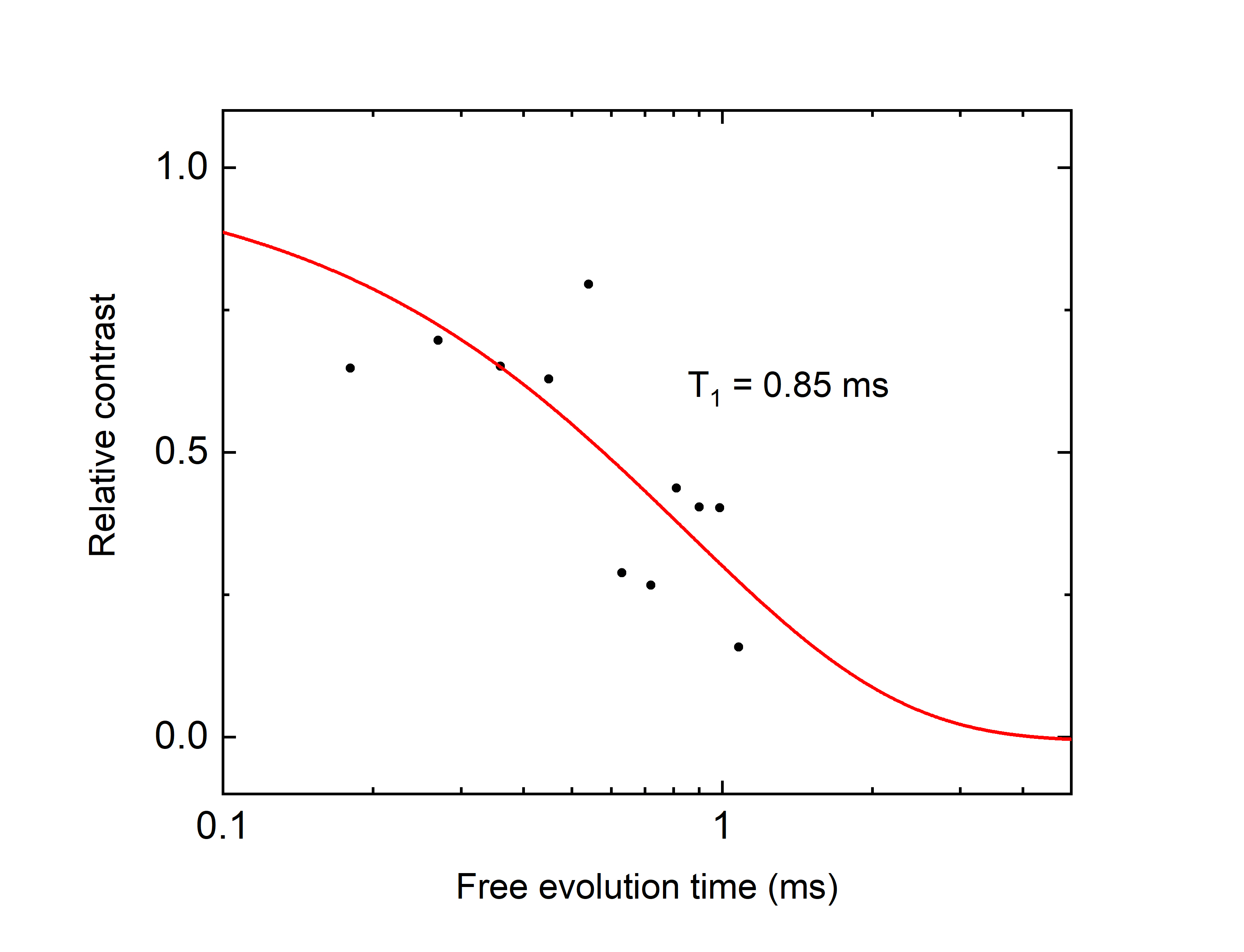}
    \caption{\textit{T$_1$} measurement done on a nanodiamond grown on a molybdenum layer. \textit{T$_1$} is just as high as for nanodiamonds grown on silicon.}
    \label{moly}
\end{figure}

Due to the use of silicon substrates, NDs usually also contain Silicon Vacancy (SiV) centres, figure \ref{SI:SiliconVacancies}. The inclusion of these defects can be minimised by choosing different substrates, such as sapphire (figure \ref{SI:SapphireGrowth}). Additionally, to scale up FND production, FNDs could be grown on thin intermediate layers which can easily be removed from the initial substrate after growth by wet etching. Metal incorporation into the diamond lattice during CVD growth is known to be extremely low, even in the case of, for example, catalysts such as Ni. \cite{ARTICLE:Wolfer2010} These metal atoms usually tend to be located at the grain boundaries, and, in the case of single crystals nanodiamonds, such as those grown in this paper, this is very limited. Experiments where metal incorporation into diamond was studied using HR-TEM have shown similar results.\cite{DEGUTIS2016163} Since large metal atoms cannot diffuse into the diamond, the only incorporation mechanism is that the atoms from the metal get into the gas phase by etching of the substrate and by this way get incorporated into the lattice. We have done a preliminary test and have deposited a Mo layer on sapphire to enable easy ND removal and reduce Si contamination. Molybdenum is used in every diamond reactor even for highest purity diamond and does not incorporate in diamond, therefore we suggest it as an intermediately layer. We have demonstrated growth of ND on molybdenum and measured T$_1$ times similar as in the experiments with Si, figure \ref{moly}.

\subsection{Photoluminescence Distributions at Different Laser Powers}

\begin{figure}[H]
\centering
\includegraphics[width=\columnwidth]{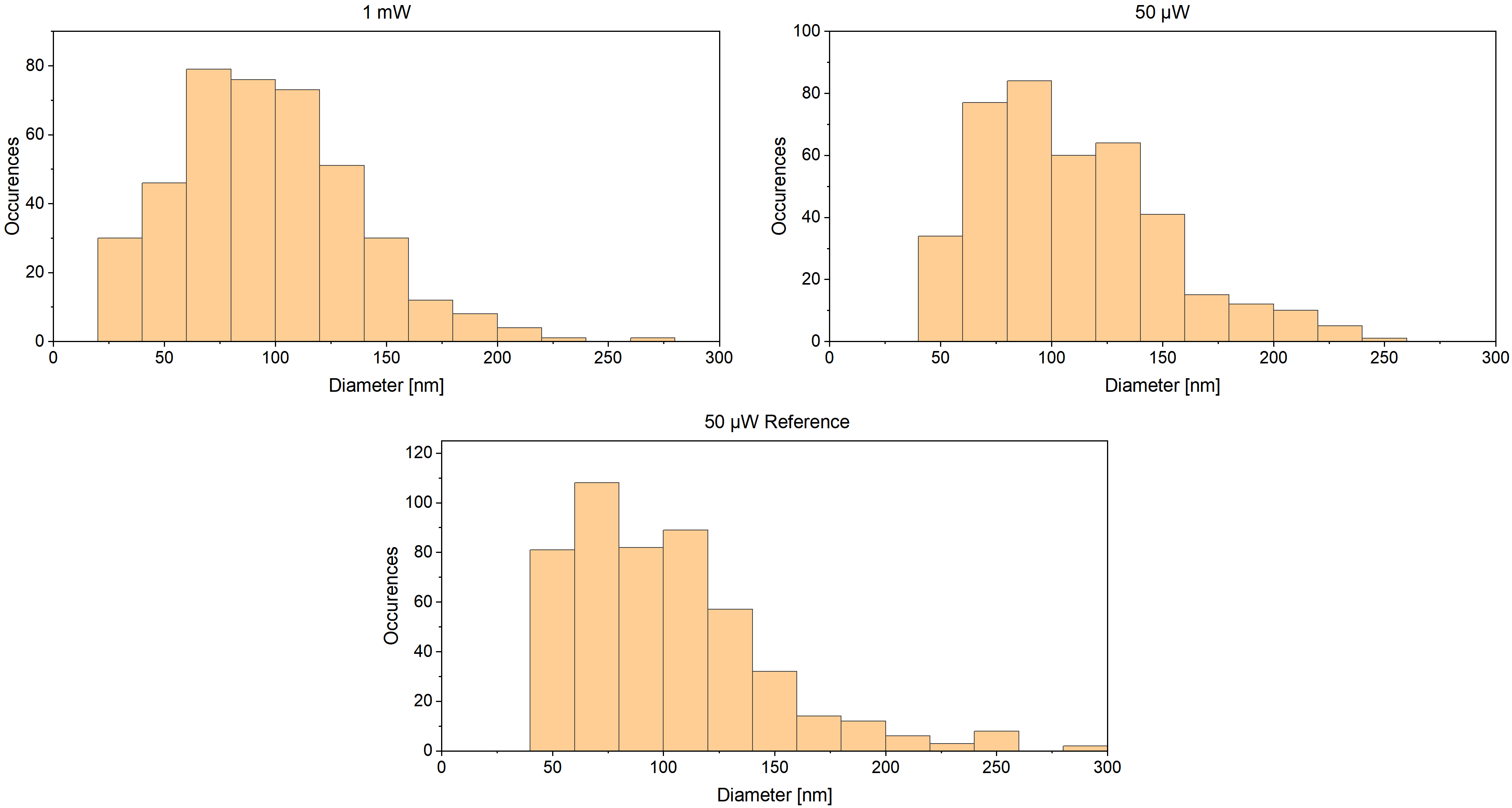}
\captionof{figure}{Calculated size distributions for PL measurements at 50 µW and 1 mW. 50 µW Reference was used to calculate the fitting parameter in table \ref{SI:PLFitParams}, and the other two distributions were calculated using these parameters. For the 1 mW distribution, single NV luminescence (I$_s$ of 22 kCounts was used. All three distributions show good consistency, with the distribution at 1 mW including smaller particles.}
\label{SI:LaserPowerDistributions}
\end{figure}

\begin{figure}[H]
\centering
\includegraphics[width=0.5\columnwidth]{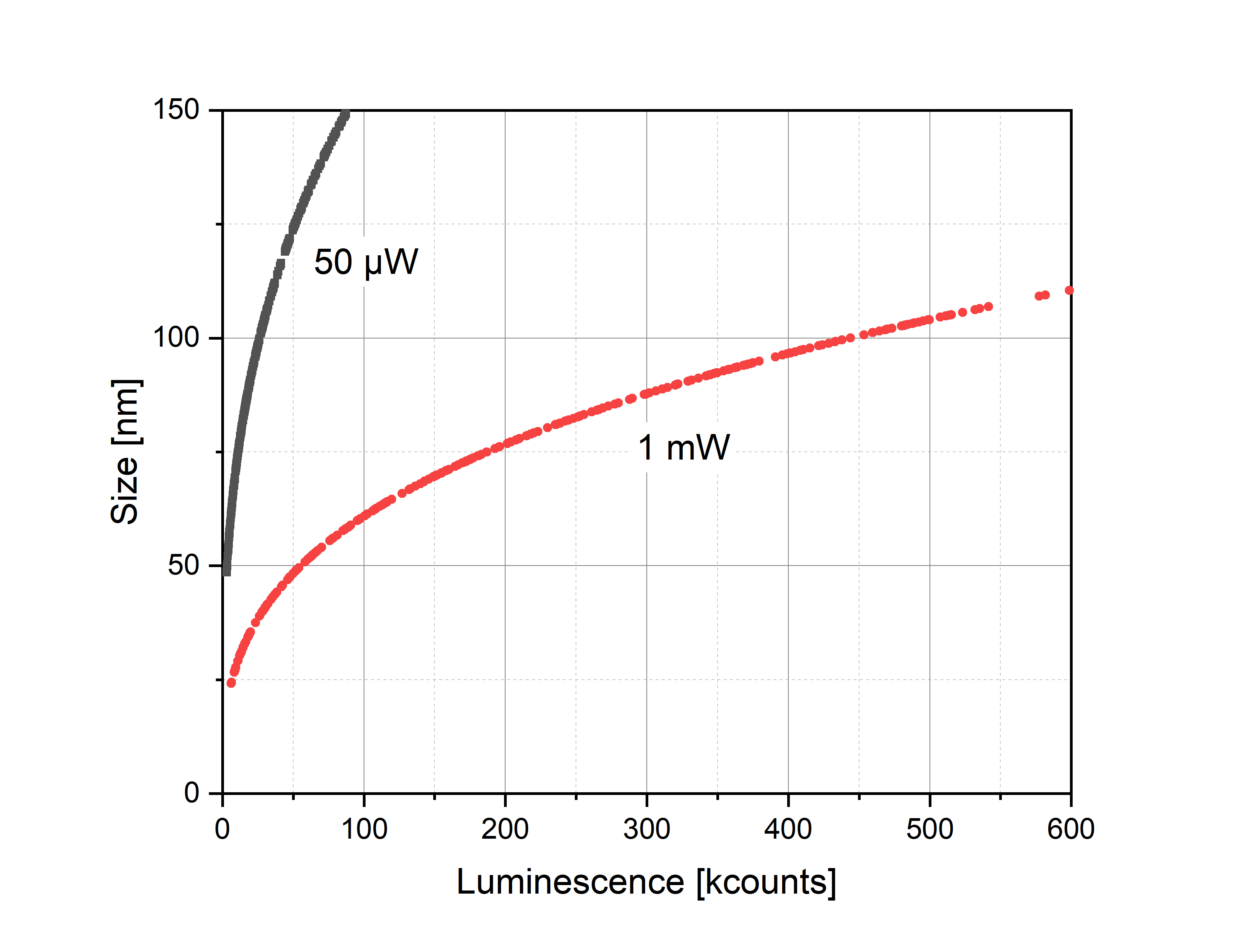}
\captionof{figure}{Size as a function of luminescence at different laser powers. At 50 µW, small particles (100 nm or less) have low photoluminescence, making them practically difficult to measure. At 1 mW, similar particles show much higher luminescence. It was assumed nitrogen was uniformly incorporated into all particles.}
\label{SI:SizeVSLum}
\end{figure}

Because of their low NV content, smaller particles blend into the background luminescence in low-power photoluminescence measurements (50 µW). To avoid this problem, some measurements are performed at a higher laser power (1mW). The calculated size distributions for particles grown at 700 °C under different laser powers are shown in figure \ref{SI:LaserPowerDistributions}. The "50 µW Reference" is the size distribution for which the corresponding fitting parameters were found in table \ref{SI:PLFitParams}. The other two distributions displayed are calculated from the same parameters, except for 1 mW a single NV luminescence of 22 kCounts was used. It demonstrates that with higher laser power, smaller particles can be found more easily. While there are 50 nm particles in the distributions measured at 50 µW, there were issues to find and measure them in practise as we couldn't track them during the measurement. To accurately follow them, we required higher illumination. This is shown in figure \ref{SI:SizeVSLum}, where we show the particle size as a function of its luminescence for 50 µW and 1 mW of laser power. At 50 µW, particles of roughly 100 nm would correspond to 26 kCounts, whereas they would correspond to 440 kCounts at 1 mW. This makes them easier to measure at higher powers, as their contrast with the background is increased. 

\begin{figure}
\centering
  \includegraphics[width=\columnwidth]{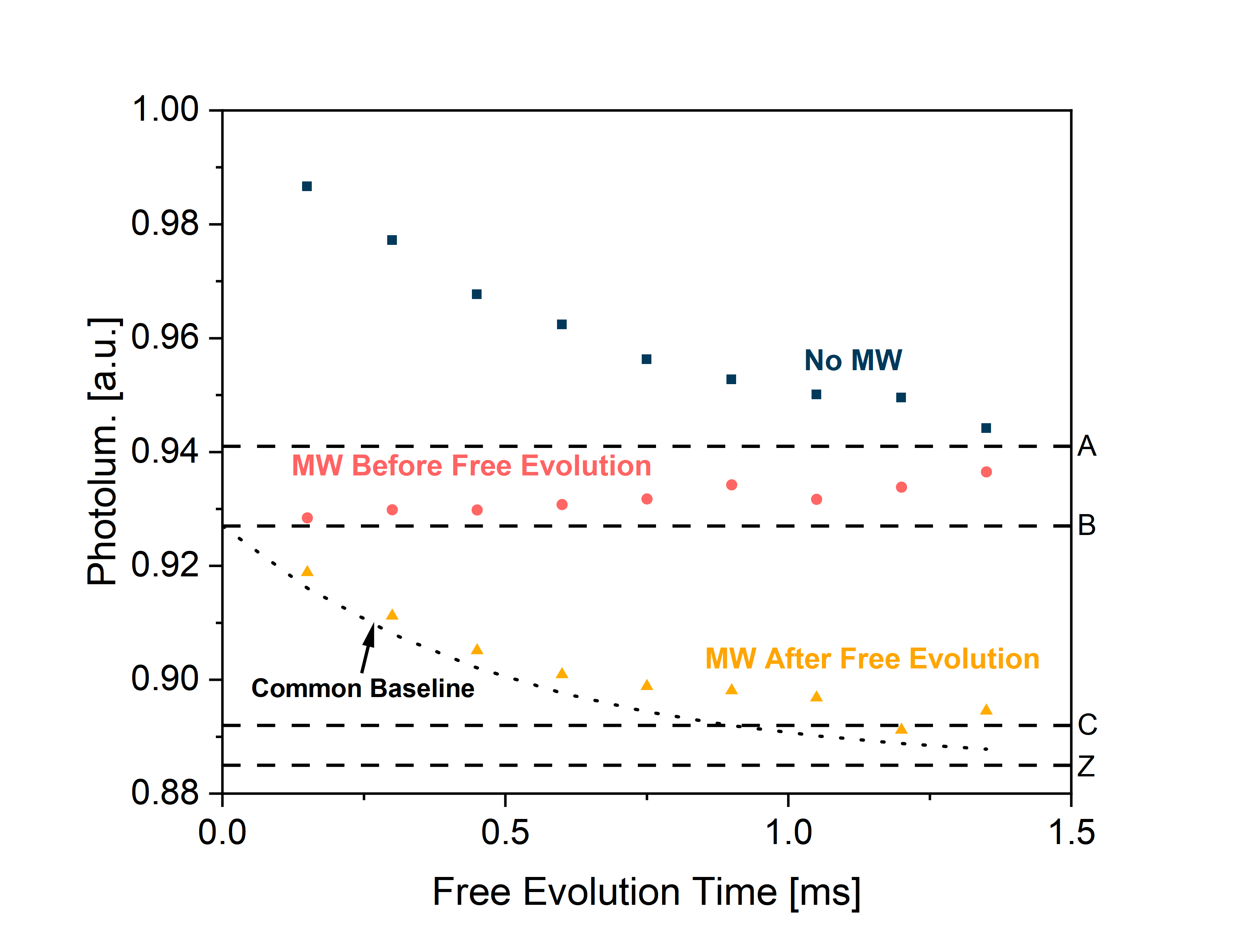}
\caption{Photoluminescence as function of free evolution time and the baseline for all three measurements. Parameters A and C are the asymptotic lines, parameter B is the starting point of the "MW" curves, the "no MW" curve starts from 1, Z is the asymptotic line for the common baseline and is calculated from A, B and C.}
\label{nd1}
\end{figure}

\begin{figure}
\centering
  \includegraphics[width=\columnwidth]{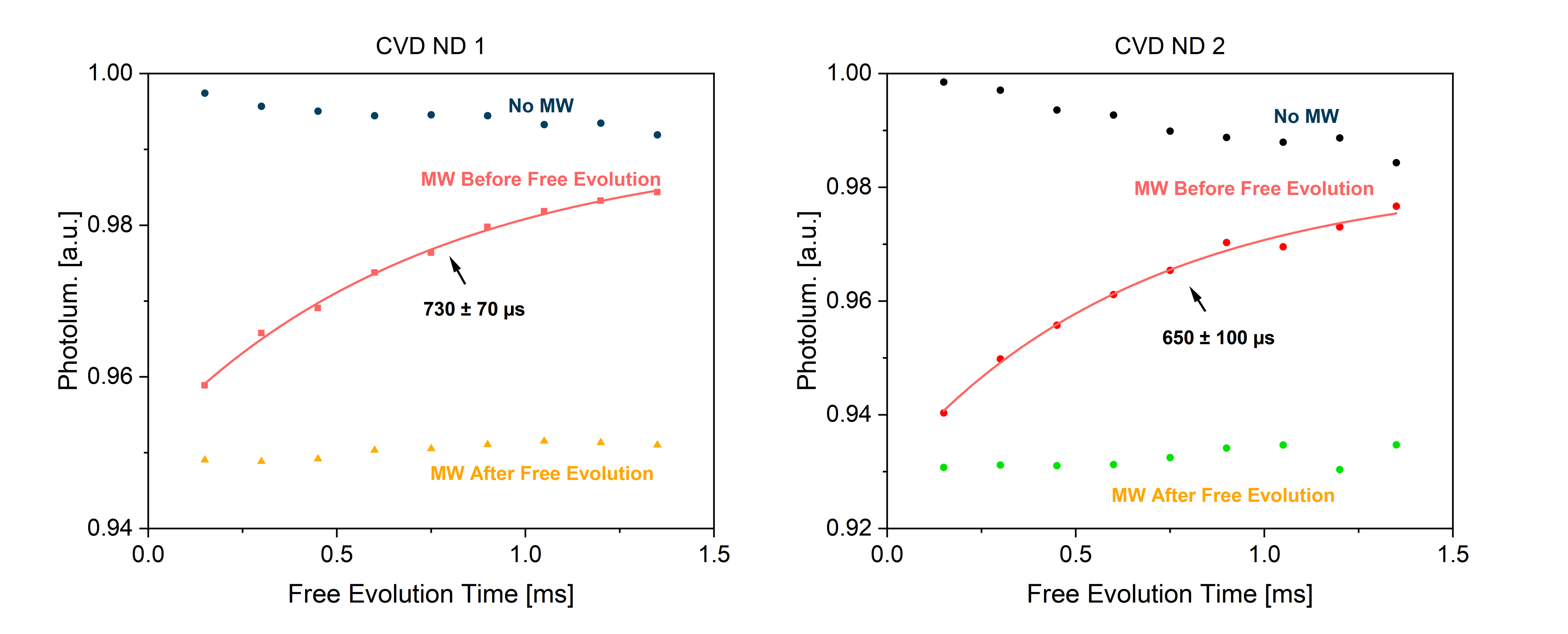}
\caption{Baseline corrected photoluminescence as function of time interval between initialisation and readout for two different shell-doped nanodiamonds of the same sample. Dark blue and yellow lines could not be fitted, therefore relaxation time \textit{T$_1$} is estimated from the red line fits.}
\label{nd2}
\end{figure}

\section{Baseline correction}

\indent The most prominent feature of the data presented in Figure 5 is the drastic difference between the "MW Before" versus "MW After" curves. In order to perform a proper analysis of the data, we develop and perform baseline correction first.\\

\indent For a perfect microwave assisted measurement, the "No MW" line in Figure \ref{nd1} should decline twice as much as the "MW After" curve rises. That is because NV photoluminescence is a linear function of state $m_s=0$ population (we know this since Rabi nutations of luminescence signal are sinusoidal) and because the $m_s=0$ state decays to both $m_s=1$ and $m_s=-1$ populations. Also, the sum of populations of all three states is constant.
\begin{equation}
\frac{\Delta L_{\text{No MW}}}{\Delta L_{\text{MW After}}}=\frac{\Delta P_{0}}{\Delta P_{+1}}
\end{equation}
\begin{equation}
P_{+1}=P_{-1}
\end{equation}
\begin{equation}
P_0+P_{+1}+P_{-1}=1
\end{equation}
where L is the photoluminescence signal and P is the population of the corresponding spin state and $\Delta$P is the change in the population at a given time from the start of the decay. This leads to
\begin{equation}
\Delta L_{\text{No MW}}=-2\Delta L_{\text{MW After}}
\label{cond}
\end{equation}
The "No MW" signal decreases therefore twice as much as the "MW After" signal increases. We find our common baseline based on the assumption that, after baseline correction, the condition expressed in equation \ref{cond} would be satisfied.\\
\indent In Figure \ref{nd1}, we re-plot the experimental data together with asymptotic lines discussed in the model below, representing the theoretical relaxed state PL value and the relaxed state including an additional not spin related PL decay. We see that the end point of the "MW After" curve would be corrected and therefore increase by a value $C-Z$, where Z is the calculated common asymptotic baseline value and C is the value of the measured PL decay for "MW After" signal.  The "No MW" curve has to then show a decrease of $2(C-Z)$, this value should equal $(1-B)-(A-Z)$. Based on equations (6), (7), (8)  we can predict the  $Z$ value.

\begin{equation}
2(C-Z)=(1-B)-(A-Z)
\end{equation}
\begin{equation}
Z=\frac{A+B+2C-1}{3}
\end{equation}
Lastly we assume that the baseline is an exponential decay. The equation for it is then
\begin{equation}
y=(B-Z)\exp{-\frac{t}{T_1'}}+Z
\end{equation}
$T_1'$ is estimated approximately from the shape of the experimental curves. With this, we are able to apply baseline correction to the data (Figure \ref{nd2}).\\

\indent We see that after baseline correction, the "No MW" line shows just a very slight decrease compared to the effect of microwave resonant pulse. This means that the population of the $m_s=0$ state changes much more significantly as a result of the population swap between the $m_s=0$ and $m_s=1$ states than from relaxation. There are two phenomena that affect the relative magnitude of these changes: imperfect initialisation - $P_0(t=0)\neq 1$, and imperfect thermalisation - $P_0(t\rightarrow\infty)\neq 0.33$. A possible reason for the former could be inhomogeneous laser excitation throughout the crystal; for the latter, leakage of laser light during free evolution.\\
\indent The leakage of laser light in our setup is, however, rather small - only $10^{-4}$, while free evolution time exceeds initialisation time only 100 times in our experiments. Also significant leakage would introduce a decrease in photoluminescence decay rate.
\begin{equation}
    \dot X(t)=[X_\text{thermal}-X(t)]R+[1-X(t)]K
\end{equation}
where $X\equiv P_0$, $R\equiv \frac1{T_1}$ and $K$ is the rate of population flow due to leakage of laser light.
\begin{equation}
    X(t)=\left(1-\frac{X_\text{thermal}R+K}{R+K}\right)\exp{\left[-\left(R+K\right)t\right]}+\frac{X_\text{thermal}R+K}{R+K}
\end{equation}
Most of our $T_1$ measurements are performed at 50 µW laser power, but some at 1 mW. We do not observe a significant difference in photoluminescence decay. 
\begin{equation}
    \exp{\left[-\left(R+K\right)t\right]}\approx\exp{\left[-\left(R+20K\right)t\right]}
\end{equation}
This is an evidence that $K\ll R$, and this means that the leakage does not effect the population of spin states significantly.
\begin{equation}
    X(t\rightarrow\infty)=\frac{X_\text{thermal}R+K}{R+K}
\end{equation}
From this we conclude that leakage of laser light is unlikely to be the reason for such a large change in state 0 population after relaxation.\\
\indent The reason for the observed baseline drift could be low MW driving efficiency, however, the full resolution of 8 ODMR peaks and Rabi measurements for each NV orientation did not improve the inversion strength, so we attribute the differences to the strong charge effects accompanying the NV excitation and photoionisation even for very low powers used.\par
It is not necessary to find a common baseline equation for every measurement; a subtraction of two curves from Figure \ref{nd1} already reveals the baseline corrected relaxation time; however, finding the exact shape of the baseline is important in understanding its origin.

\section{\textit{T$_2$} measurements}

\begin{figure}[H]
    \centering
    \includegraphics[width=0.6\columnwidth]{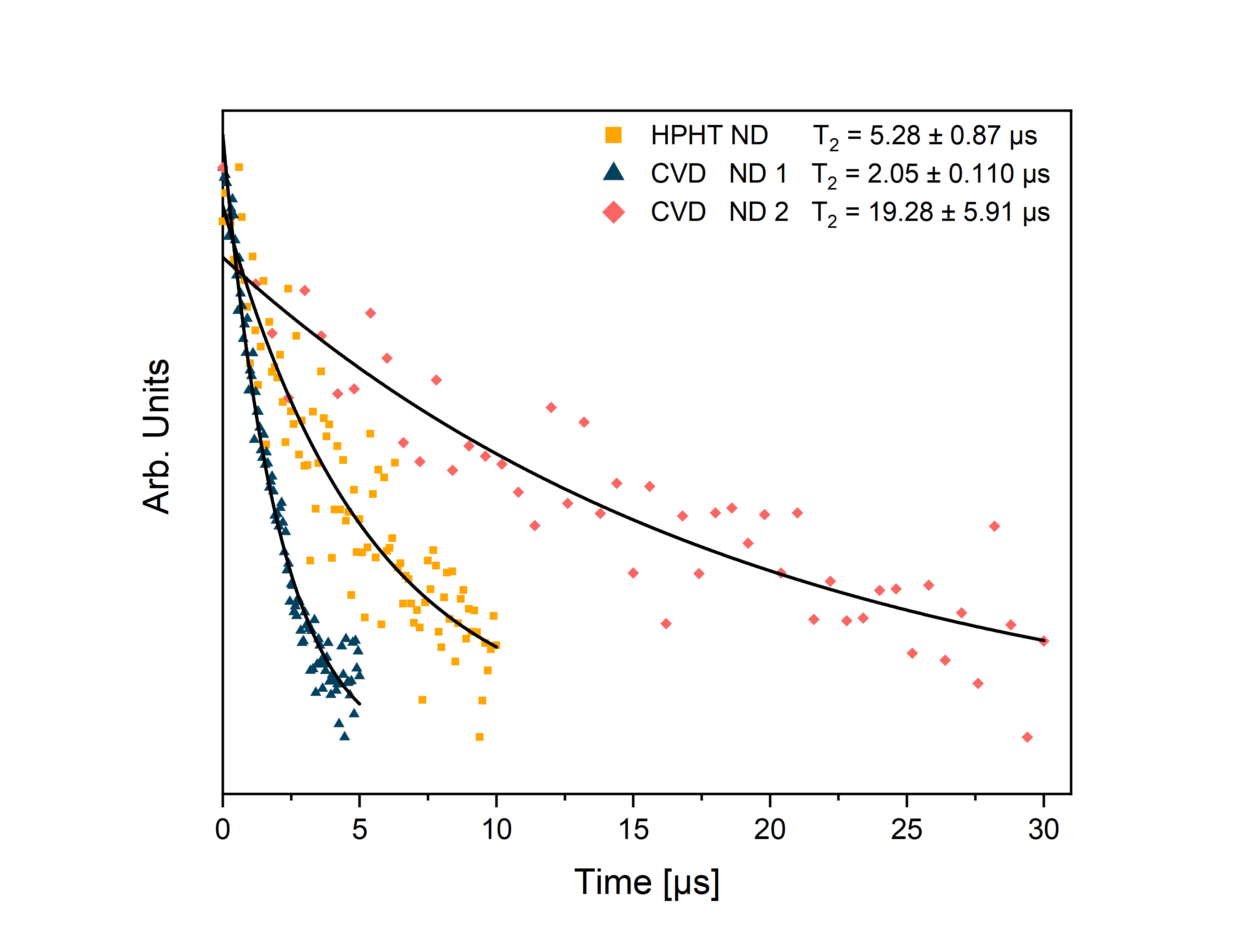}
    \captionof{figure}{\textit{T$_2$} measurements done on reference HPHT NDs sample, and on two shell-doped CVD NDs. \textit{T$_2$} measurements do not show similar improvements as those achieved with \textit{T$_1$} relaxation.}
    \label{fig:T2Meas}
\end{figure}

Transverse relaxation times were also measured with Hahn-echo sequences on some of the shell-doped CVD and HPHT particles, figure \ref{fig:T2Meas}. We have found that the $T_2$ time for our doped samples is in the order of 1 µs. This is significantly lower than in bulk samples \cite{Herbschleb2019}. The reason for such short coherence time could be presence of strong magnetic noise on a similar time scale as the spin coherence time. 
These values are, however, rather typical for NV ensembles \cite{Rondin_2014}. In fact, for single NV nanodiamonds it is possible to achieve \textit{T$_2$} even up to 400 µs with dynamical decoupling.\cite{PhysRevB.105.205401} Due to the fact that the charges and radicals present at the ND surface can provide a strong $E$ and $B$ field bath, low coherence times are within expectations. However, the \textit{T$_1$} time stays high in such samples as expected to the mechanism of \textit{T$_1$} relaxation.

\section{Relaxation from different spin states}µ

Even though the relaxation path from $m_s=1$ is different than from $m_s=0$ state, the relaxation time is the same. This is supported by rate equations. Since transition rates are the same for the symmetric transitions within the ground state triplet
\begin{equation}
    k_{0\longrightarrow1}=k_{0\longrightarrow-1}=k_0
\end{equation}
\begin{equation}
    k_{1\longrightarrow0}=k_{-1\longrightarrow0}=k_1
\end{equation}
we can write the population flux into $m_s=0$ state the following way:
\begin{equation}
    F_{\longrightarrow0}=\frac{\text dn_0}{\text dt}=(n_1+n_{-1})k_1-2n_0k_0
\end{equation}
where $n$ is the population of the respective state. Since $n_0+n_1+n_{-1}=1$, we get
\begin{equation}
    \frac{\text dn_0}{\text dt}=-(k_1+2k_0)n_0+k_1
\end{equation}
\begin{equation}
    n_0=Ce^{-(k_1+2k_0)t}+\frac{k_1}{k_1+2k_0}
\end{equation}
where $C$ is the constant that depends on initial conditions.
\begin{equation}
    T_1=\frac 1{k_1+2k_0}
\end{equation}
Hence $T_1$ is independent of initial conditions.

\bibliographystyle{plain}
\bibliography{papers.bib}